\documentclass[aps,pra,twocolumn,groupedaddress,floatfix]{revtex4-1}

\usepackage{amsmath}
\usepackage{amssymb}
\usepackage{mathtools}
\usepackage{multirow}
\usepackage{float}
\usepackage[range-phrase=--, range-units = single,retain-explicit-plus]{siunitx}
\usepackage{graphicx}
\usepackage{braket}
\usepackage{hyperref}
\usepackage{color}

\newcommand{\tsup}{\textsuperscript}	
\newcommand{\tsub}{\textsubscript}		
\newcommand{\Sr}[1]{\tsup{#1}\textnormal{Sr}}		
\newcommand{\Srion}[1]{\textnormal{Sr}\tsup{#1}}	
\newcommand{\Li}[1]{\tsup{#1}\textnormal{Li}}		
\newcommand{\SLJ}[3]{\tsup{#1}\textnormal{#2}\tsub{#3}}
\newcommand{\SLJF}[4]{\tsup{#1}\textnormal{#2}\tsub{#3}\textnormal{, }{F = {#4}}}
\newcommand{\nSLJ}[4]{{#1}\textnormal{ }\SLJ{#2}{#3}{#4}}

\newcommand{\nSLJF}[5]{{#1}\textnormal{ }\SLJ{#2}{#3}{#4}\textnormal{, }{F = {#5}}}

\newcommand{\solidrule}[1][1cm]{\rule[0.3ex]{#1}{2pt}}
\newcommand{\solidline}{\mbox{\solidrule[7mm]}}
\newcommand{\dashedline}{\mbox{\solidrule[2mm]\hspace{1mm}\solidrule[2mm]\hspace{1mm}\solidrule[2mm]}}
\newcommand{\dottedline}{\mbox{\solidrule[0.7mm]\hspace{1.4mm}\solidrule[0.7mm]\hspace{1.4mm}\solidrule[0.7mm]\hspace{1.4mm}\solidrule[0.7mm]}}
\newcommand{\dotdashedline}{\mbox{\solidrule[2.25mm]\hspace{1mm}\solidrule[.5mm]\hspace{1mm}\solidrule[2.26mm]}}

\DeclareSIUnit\atoms{atoms}

\begin{document}


\title{Spectroscopy of \Sr{87} triplet Rydberg states}

\author{R. Ding}
\author{J. D. Whalen}
\author{S. K. Kanungo}
\author{T. C. Killian}
\author{F. B. Dunning}
\affiliation{Department of Physics and Astronomy, Rice University, Houston, Texas 77005, USA}

\author{S. Yoshida}
\author{J. Burgd\"{o}rfer}
\affiliation{Institute for Theoretical Physics, Vienna University of Technology, A-1040 Vienna, Austria, EU}

\date{\today}

\begin{abstract}
A combined experimental and theoretical spectroscopic study of {high-$n$}, ${30 \lesssim n \lesssim 100}$, triplet $\SLJ{}{S}{}$ and $\SLJ{}{D}{}$ Rydberg states in \Sr{87} is presented. 
\Sr{87} has a large nuclear spin, ${I=9/2}$, and at {high-$n$} the hyperfine interaction becomes comparable to, or even larger than, the fine structure and singlet-triplet splittings which poses a considerable challenge both for precision spectroscopy and for theory.
For {high-$n$} $\SLJ{}{S}{}$ states, the hyperfine shifts are evaluated non-perturbatively taking advantage of earlier spectroscopic data for the ${I=0}$ isotope \Sr{88}, which results in good agreement with the present measurements. 
For the $\SLJ{}{D}{}$ states, this procedure is reversed by first extracting from the present \Sr{87} measurements the energies of the $\SLJ{3}{D}{1,2,3}$ states to be expected for isotopes without hyperfine structure (\Sr{88}) which allows the determination of corrected quantum defects in the {high-$n$} limit.
\end{abstract}


\maketitle


\section{Introduction}
\label{S:into}

Rydberg excitation in dense cold atom samples can lead to the formation of ultralong-range Rydberg molecules in which scattering of the Rydberg electron from neighboring ground-state atoms leads to the binding of one, or more, ground-state atoms in multiple possible vibrational levels \cite{gds00, bbn09, lpr11, trb12, dad15, bcb13, smd15, kgb14, amr14, bry15, egr15, elp17, gkb14, dky16}. 
Measurements of such weakly-bound Rydberg molecules have also been extended to dense BECs and higher $n$ values where the Rydberg electron orbit can enclose tens to hundreds of ground-state atoms \cite{bbn11, csw18, ked18}.

The interaction between the excited Rydberg electron and a ground-state atom can be described using a Fermi pseudopotential. 
For strontium, except at short ranges, {$s$-wave} scattering dominates due to the lack of a {$p$-wave} resonance. 
This results in an oscillating molecular potential that reflects the modulations in the electron probability density \cite{bbn09}. 
The largest, and deepest, potential well is located near the outer classical turning point and the wave function of the ground vibrational state of the Rydberg molecule is strongly localized in this region. 
Thus, the probability for forming a ground-state dimer molecule will depend on the likelihood of initially finding a pair of ground-states atoms at the appropriate internuclear separation, $R$. 
By varying $n$, and the location of the potential minimum, one can probe the pair correlation function in the ultracold gas. 
This provides an opportunity to examine the influence of quantum statistical properties on Rydberg molecule formation. 
Strontium is an attractive candidate for such a study because it possesses both bosonic (\Sr{84}, \Sr{86}, \Sr{88}) and fermionic (\Sr{87}) isotopes, all of which have been cooled to degeneracy. 
The excitation spectra for the bosonic isotopes are particularly simple as they have zero nuclear spin (${I=0}$) and therefore no hyperfine structure. 
In contrast, \Sr{87} has nuclear spin ${I=9/2}$ which results in hyperfine interactions that greatly complicate the excitation spectrum.  

Several studies of Rydberg spectra for bosonic \Sr{88} have been reported \cite{vail12, eshe77, beig82b}. 
These studies primarily centered on lower $n$ states (${n \lesssim 40}$) and focused on the perturbations introduced by channel interactions and their treatment using multichannel quantum defect theory (MQDT). 
Information on {higher-$n$} levels was, typically, obtained by extrapolating the measured quantum defects using the Rydberg-Ritz formula. 
Such extrapolation is known to be an effective method for predicting the energies of {high-$n$} Rydberg states whose quantum defects are essentially {$n$-independent} and therefore nearly constant. 
This, however, is not true for strontium $\SLJ{}{D}{}$ states whose quantum defects exhibit a relatively strong {$n$-dependence}. 

Experimental and theoretical studies of the spectrum for \Sr{87} have also been reported \cite{beig81, beig81errata, beig82a, beig83, beig88, sun89}. 
These include measurements at low $n$ where the hyperfine interaction can be treated as a weak perturbation, and at {high-$n$} (${n \sim 100}$) where the hyperfine shift becomes comparable to, or even larger than, the energy spacing between adjacent unperturbed states. 
Analysis of the {high-$n$} spectrum, therefore, poses a considerable challenge and requires use of non-perturbative methods. 
One possible approach is to take advantage of the accurate spectral information available for the bosonic isotope \Sr{88} and use it to estimate the spectrum for \Sr{87} \cite{beig81, beig81errata, beig82a, beig83}. 
For $\SLJ{}{S}{}$-states this approach provides energy levels that agree reasonably well with measured data \cite{beig81, beig81errata, beig82a, beig83}. 
A similar method utilizing a truncated basis set has been used to study {low-$n$} (${n < 20}$) \Sr{87} $\SLJ{}{D}{}$-states \cite{beig82c}. 
However, the {high-$n$} levels were analyzed by MQDT \cite{sun89} because no corresponding measured levels for the bosonic isotopes were available. 
Earlier spectroscopic studies utilized a heat pipe which can introduce uncertainties due to Doppler and pressure broadening. 
Moreover, Stark shifts due to the presence of stray fields could not be controlled. 
Indeed, for {high-$n$} states, ${n \gtrsim 100}$, additional ad-hoc corrections were introduced to obtain agreement between the theoretical estimates and the experimental measurements. 

In this work, we have measured and analyzed the excitation spectrum for {high-$n$} (${50\lesssim n \lesssim 100}$) $\SLJ{}{S}{}$ and $\SLJ{}{D}{}$ Rydberg states created in an \Sr{87} ultracold gas using two-photon excitation as a precursor to planned studies of Rydberg molecule formation in fermionic gases. 
Measurements using ultracold atoms are expected to be more accurate than measurements in a heat pipe because Doppler and pressure broadening are well suppressed and stray fields can also be controlled. 
In the present two-photon excitation scheme the intermediate $\nSLJ{5s5p}{3}{P}{1}$ state is used instead of the $\nSLJ{5s5p}{1}{P}{1}$ state employed in earlier studies. 
Since the $\nSLJ{5s5p}{3}{P}{1}$ state has a much longer lifetime than the $\nSLJ{5s5p}{1}{P}{1}$ state ({$\Gamma/2\pi = \SI{7.5}{\kHz}$} and {$\Gamma/2\pi = \SI{32}{\MHz}$}, respectively), broadening induced by scattering off the intermediate state is also suppressed. 

We compare our experimental data with predictions derived from a semi-empirical theoretical description that exploits spectroscopic data for the bosonic isotopes. 
This approach produces satisfactory agreement with the present measurements. 
We also derive improved Rydberg-Ritz formulae for both $\SLJ{}{S}{}$ and $\SLJ{}{D}{}$ states at very {high-$n$}. 

\section{Theoretical approach}
\label{S:Theoretical approach}
 
An ab-initio theoretical description of the electronic structure of strontium Rydberg atoms with a precision of $\SI{\sim 10}{\MHz}$ or better is currently out of reach. 
Thus, in order to arrive at a quantitative and predictive description, it is necessary to resort to semi-empirical methods. 
The theoretical approach adopted here follows that of earlier work by Beigang and coworkers \cite{beig82a, beig83}.

The underlying idea is to exploit the much simpler (and for $\SLJ{}{S}{}$ states, better known) electronic structure of the bosonic isotope \Sr{88} as reference for \Sr{87} to accurately account for the perturbations introduced by hyperfine interactions by direct diagonalization. 
The spectroscopic data for \Sr{88} thus serve as an ``analogue simulation'' of the full $N$-electron Schr{\"o}dinger equation that accounts for electron correlation and configuration interactions, which are tacitly assumed to be the same for all the isotopes. 
Isotope-specific interactions are then taken into account non-perturbatively by diagonalizing the full Hamiltonian which includes the hyperfine interaction. 
Accordingly, the Hamiltonian ${H(87)}$ for \Sr{87} is written as 
\begin{equation}
H(87) = H_0(88,m_{87}) + V_{\rm HF}
\label{eq:hamil}
\end{equation}
where ${H_0(88,m_{87})}$ plays the role of the ``unperturbed'' Hamiltonian that yields the eigenstates and eigenenergies, i.e., spectral lines, for \Sr{88} but rescaled by the isotope shift corresponding to the reduced mass ${m_{87} = m_e M_{87}/(m_e + M_{87})}$ where $m_e$ is the electron mass, $M_{87}$ is the mass of {\Sr{87}\tsup{+}} ion, and $V_\mathrm{HF}$ is the hyperfine interaction. 
Corrections beyond the elementary isotope shift, in particular, the mass polarization correction, can be estimated from earlier data for helium Rydberg states \cite{cok79, cok81, veld90} and, upon re-scaling to Sr, are found to be $\SI[input-comparators=\lesssim]{\lesssim 1}{\MHz}$ and can therefore be neglected. 

The Hamiltonian ${H(87)}$ [Eq.~(\ref{eq:hamil})] is diagonalized using the basis states $\ket{((5sn\ell) \, ^{2S+1}L_J, I) F}$ constructed by the coupling of angular momenta ${\vec{F} = \vec{J} + \vec{I}}$ where $\vec{I}$ is the nuclear spin and $\ket{(5sn\ell) \, ^{2S+1}L_J}$ are the eigenstates of ${H_0(88,m_{87})}$. 
We note that we retain the conventional Russell-Saunders ${^{2S+1}L_J}$ notation for the eigenstates of ${H_0(88,m_{87})}$ even though $S$ and $L$ are not exactly conserved quantum numbers in the presence of the spin-orbit interaction. 
In this basis ${H_0(88,m_{87})}$ is diagonal with corresponding eigenenergies
\begin{equation}
E_{n,S,L,J}^{(0)} = E_{\rm ion}^{(0)} - \frac{R(m_{87})}{(n - \mu^{(0)}_{n,S,L,J})^2}
\label{eq:ene0}
\end{equation}
where $E_{\rm ion}^{(0)}$ is the energy corresponding to the first ionization threshold of \Sr{87} assuming ${I=0}$, $\mu^{(0)}_{n,S,L,J}$ is the quantum defect for the state $\ket{(5sn\ell) \, ^{2S+1}L_J}$, and ${R(m_{87}) = R_{\infty} m_{87} / m_e}$ with the Rydberg constant $R_{\infty}$. 
In the following we use either directly measured or extrapolated (at {high-$n$}) quantum defects for \Sr{88} as input.

\begin{figure}[!htbp]
\begin{center}
\includegraphics[width=8.6cm, keepaspectratio=true]{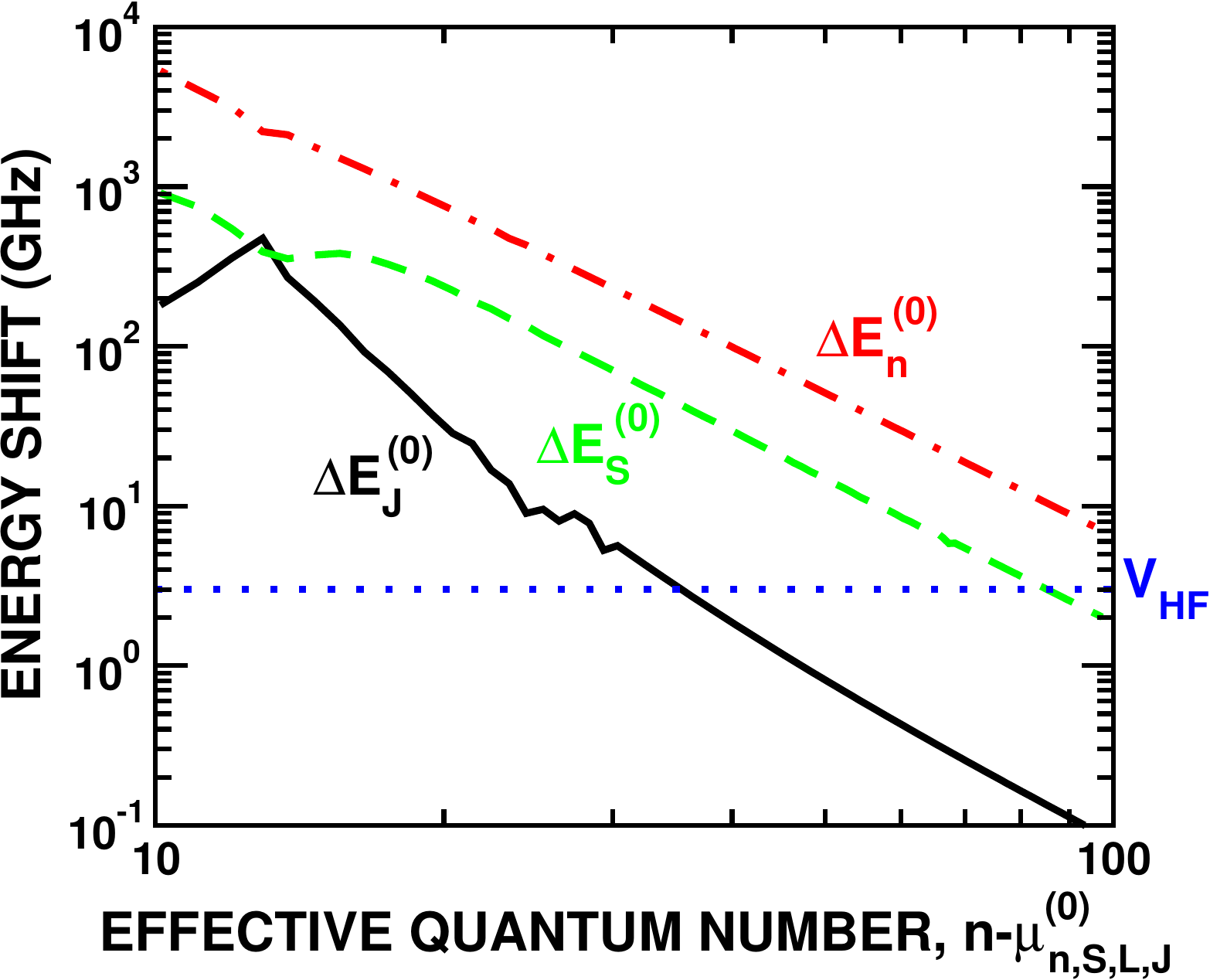}
\end{center}
\caption{\label{fig:deltaE}(Color online.) ({\color{black}\solidline}) The $n$ scaling of the fine structure splitting $\Delta E_J^{(0)}$, ({\color{green}\dashedline}) the spin singlet-triplet splitting $\Delta E_S^{(0)}$, and ({\color{red}\dotdashedline}) the level separation $\Delta E_n^{(0)}$ between like states in \Sr{88} that differ in $n$ by one. ({\color{blue}\dottedline}) The strength of the hyperfine interaction in \Sr{87}. The splittings $\Delta E_J^{(0)}$ and $V_{\rm HF}$ refer to $\SLJ{3}{D}{}$ states with ${J=1}$ and ${J=2}$ and are evaluated using the measured data and their extrapolation.}
\end{figure}

The hyperfine interaction results from the interaction between an electron and the electric and magnetic multipoles of the nucleus \cite{schw55}. 
For singly-excited {high-$n$} strontium atoms with two electrons outside closed shells, $V_{\rm HF}$ is governed by the interaction of the $5s$ valence and $n\ell$ Rydberg electrons with the \Sr{87} nuclear spin ${I=9/2}$. 
Because of the ${(n^{*})^{-3}}$ scaling of the hyperfine interaction \cite{gued93}, the hyperfine shift associated with the Rydberg electron for {high-$n$} values (${n > 20}$) can be estimated to be $\SI[input-comparators=\lesssim]{\lesssim1}{\MHz}$ and can therefore be safely neglected. 
(${n^* = n - \mu^{(0)}_{n,S,L,J}}$ is the effective quantum number and ${n^* \simeq 1.5}$ for the $\nSLJ{5s^2}{1}{S}{0}$ ground state.) 
Therefore, the hyperfine interaction $V_{\rm HF}$ can be approximated by the contact interaction of the inner (or valence) $5s$ electron with the nucleus \cite{beig83} 
\begin{equation}
V_{\rm HF} \simeq a_{5s} \vec{s}_\mathrm{in} \cdot \vec{I} \, ,
\label{eq:hf}
\end{equation}
where $\vec{s}_\mathrm{in}$ is the spin of the inner $5s$ electron. 
The hyperfine coupling constant can be extracted from the ionization limit yielding ${a_{5s} \simeq \SI{-1.0005}{\GHz}}$ \cite{suna93} [see discussion following Eq.~(\ref{eq:eion})]. 
Since the interaction of the Rydberg electron and the nuclear spin is negligibly small, the hyperfine interaction $V_{\rm HF}$ is approximately independent of $n$. 
This $n$ independence of $V_{\rm HF}$ [Eq.~(\ref{eq:hf})] has profound consequences for the Rydberg spectrum described by the isotope-rescaled Hamiltonian $H(87)$ [Eq.~(\ref{eq:hamil})]. 
The matrix elements of the reference Hamiltonian $H_0(88,m_{87})$ depend on the fine structure splitting ${\Delta E_J^{(0)} = | E_{n,S,L,J+1}^{(0)} -  E_{n,S,L,J}^{(0)} |}$ which, taking $\SLJ{}{D}{}$ states as an example, scales as
\begin{equation}
\Delta E_J^{(0)} \sim 4.4 \times 10^5/n^{*\, 3.4} \quad \mbox{(GHz)} \, .
\end{equation}
The singlet-triplet splittings scale as
\begin{equation}
\Delta E_S^{(0)} = | E_{n,1,L,J}^{(0)} -  E_{n,0,L,J}^{(0)} | \sim 1.8 \times 10^6/n^{*\, 3} \quad \mbox{(GHz)} \, ,
\end{equation}
and the Coulomb splittings scale as
\begin{equation}
\Delta E_n^{(0)} = | E_{n+1,S,L,J}^{(0)} -  E_{n,S,L,J}^{(0)} | \sim 5.8 \times 10^6/n^{*\, 3} \quad \mbox{(GHz)} \, .
\end{equation}
Therefore, as $n^*$ increases, $V_{\rm HF}$ becomes comparable in size to the fine structure splitting, the singlet-triplet splitting, and finally the Coulomb splitting.
This is illustrated in Fig.~\ref{fig:deltaE} and leads to strong state mixing. 
In consequence, Eq.~(\ref{eq:hamil}) cannot, in general, be treated perturbatively but rather must be diagonalized. 

The present approach is a variant of MQDT \cite{sun89, robi18} commonly used to analyze the energy levels of multi-electron systems. 
In MQDT, instead of describing microscopically the core-electron interaction in each channel and the mixing of different channels, interactions are represented by a set of parameters (e.g., scattering phase shifts and {$K$-matrices}) which are typically extracted from the measured data. 
In the current approach, a different set of parameters, i.e., the measured quantum defects (or equivalently, energy levels [Eq.~(\ref{eq:ene0})]) of isotopes with vanishing nuclear spin are used. 

An alternative approach to describe the energy levels in strontium is to use a two-active electron (TAE) model \cite{fiel18} which treats the electron-electron interactions between the outer electrons microscopically while their interaction with the ${N-2}$ electron core is parameterized in terms of model potentials. 
The currently available model potentials yield quantum defects with an accuracy of \num{\sim 0.01}. 
This uncertainty is larger than that present in current experimental data, especially for low $n$ states. 
Therefore, we do not employ the TAE approximation in Eq~(\ref{eq:hamil}) for deriving results to compare with experiment. 
However, we do use TAE calculations to probe the validity of the approximations entering into our semi-empirical description. 
For example, the approximation of the hyperfine interaction by the contact term [Eq.~(\ref{eq:hf})] is confirmed by TAE calculations. 
Contributions from the interactions between the Rydberg electron and the magnetic dipole and electric quadrupole moments of the core ion are found to be of the order of \SI{100}{\Hz} (or smaller) around ${n=100}$. 
Moreover, the mixing of $4dn\ell$ and $5pn\ell$ channels in the $\ket{(5sn\ell) \, ^{2S+1}L_J}$ state is negligibly small (\SI{<0.02}{\percent}) and, therefore, the polarization of the second (inner) valence electron can be neglected.

In the following we consider two-photon excitation of \Sr{87} from the ground state to $\SLJ{}{S}{}$ or $\SLJ{}{D}{}$ Rydberg states. 
In the limit ${n \to \infty}$ both the $\SLJ{}{S}{}$ and $\SLJ{}{D}{}$ Rydberg states converge to the $\Srion{+}$ $(\nSLJ{5s}{2}{S}{1/2})$ ionization limit. 
Because of the hyperfine interaction, this ionization limit is split into two components with {${F=4}$ or $5$},
\begin{equation}
E_{\rm ion}(F) = E_{\rm ion}^{(0)} + \frac{a_{5s}}{2} \left(F(F+1) - I(I+1) - \frac{3}{4}\right) \, ,
\label{eq:eion}
\end{equation}
where $E_{\rm ion}^{(0)}$ is the threshold for \Sr{87} assuming its nuclear spin ${I=0}$. 
From the splitting of the ionization thresholds ${E_{\rm ion}(F=4) - E_{\rm ion}(F=5)}$, the hyperfine constant $a_{5s}$ is determined.

\subsection{Energy shift of $\SLJ{}{S}{}$ states}

In \Sr{87}, there are four $\SLJ{}{S}{}$ basis states present within a single Rydberg $n$ manifold with ${m_F = 0}$, i.e., $\ket{((5sns) \, \SLJ{1}{S}{0}, I) F=I}$ and $\ket{((5sns) \, \SLJ{3}{S}{1}, I) F=I,I\pm1}$. 
(Note that the hyperfine interaction is independent of $m_F$.) 
For evaluation of the matrix elements of the hyperfine interaction $V_{\rm HF}$ in this basis the angular integrals can be performed analytically \cite{luri62}. 
Since $F$ is an exact quantum number, substates of different $F$ remain decoupled under the action of $V_{\rm HF}$. 
Consequently, the hyperfine shifts of the states ${F = I \pm 1}$ are given by the diagonal elements of the matrix $V_{\rm HF}$
\begin{eqnarray}
&& \bra{((5sns) \, \SLJ{3}{S}{1}, I) F=I+1} V_{\rm HF} \ket{((5sns) \, \SLJ{3}{S}{1}, I) F=I+1} \nonumber \\
&& \qquad\qquad = {\frac12 a_{5s} I \simeq \SI{-2.25}{\GHz}}
\label{eq:hfs+1}
\end{eqnarray}
and
\begin{eqnarray}
&& \bra{((5sns) \, \SLJ{3}{S}{1}, I) F=I-1} V_{\rm HF} \ket{((5sns) \, \SLJ{3}{S}{1}, I) F=I-1} \nonumber \\
&& \qquad\qquad = {- \frac12 a_{5s} (I+1) \simeq \SI{2.75}{\GHz}} \, .
\label{eq:hfs-1}
\end{eqnarray}
Because of the orthogonality of the radial wavefunctions, states with different $n$ belonging to the same spin multiplet are decoupled. 
In the limit ${n \to \infty}$, these states converge to the ionization limits ${E_{\rm ion}(F=I \pm 1/2)}$ [Eq.~(\ref{eq:eion})] associated with the states $\nSLJ{5s}{2}{S}{1/2}$, ${F=5}$ [Eq.~(\ref{eq:hfs+1})] or ${F=4}$ [Eq.~(\ref{eq:hfs-1})] of the \Srion{+} ion. 
For ${F=I}$, the hyperfine interaction causes singlet-triplet mixing and leads to a breakdown of the $LS$ coupling  scheme. 
Since the radial functions belonging to different spin multiplets are not pairwise orthogonal, the matrix $V_{\rm HF}$ for the the subspace ${F=I}$ becomes
\begin{eqnarray}
&& \bra{((5sn's) \, \SLJ{1}{S}{0}, I) F=I} V_{\rm HF} \ket{((5sns) \, \SLJ{1}{S}{0}, I) F=I} \nonumber \\
&& \qquad\qquad = 0 \label{eq:hfs1} \\
&& \bra{((5sn's) \, \SLJ{3}{S}{1}, I) F=I} V_{\rm HF} \ket{((5sns) \, \SLJ{3}{S}{1}, I) F=I} \nonumber \\
&& \qquad\qquad = - \frac{1}{2} a_{5s} \delta_{n,n'} \label{eq:hfs2} \\
&& \bra{((5sn's) \, \SLJ{1}{S}{0}, I) F=I} V_{\rm HF} \ket{((5sns) \, \SLJ{3}{S}{1}, I) F=I} \nonumber \\
&& \qquad\qquad = \frac{1}{2} a_{5s} \sqrt{I(I+1)} O_{n,n'} \, , \label{eq:hfs3}
\end{eqnarray}
where $O_{n,n'}$ is the overlap between the singlet and the triplet radial wavefunctions and can be estimated semiclassically \cite{bhat81}. 
For example, ${O_{n,n'} \simeq 0.98}$ for ${n=n'}$, \num[input-comparators=\simeq]{\simeq 0.1} for ${|n-n'|=1}$, and continues to rapidly decrease with increasing ${|n-n'|}$. 

\begin{figure}[!htbp]
\begin{center}
\includegraphics[width=8.6cm, keepaspectratio=true]{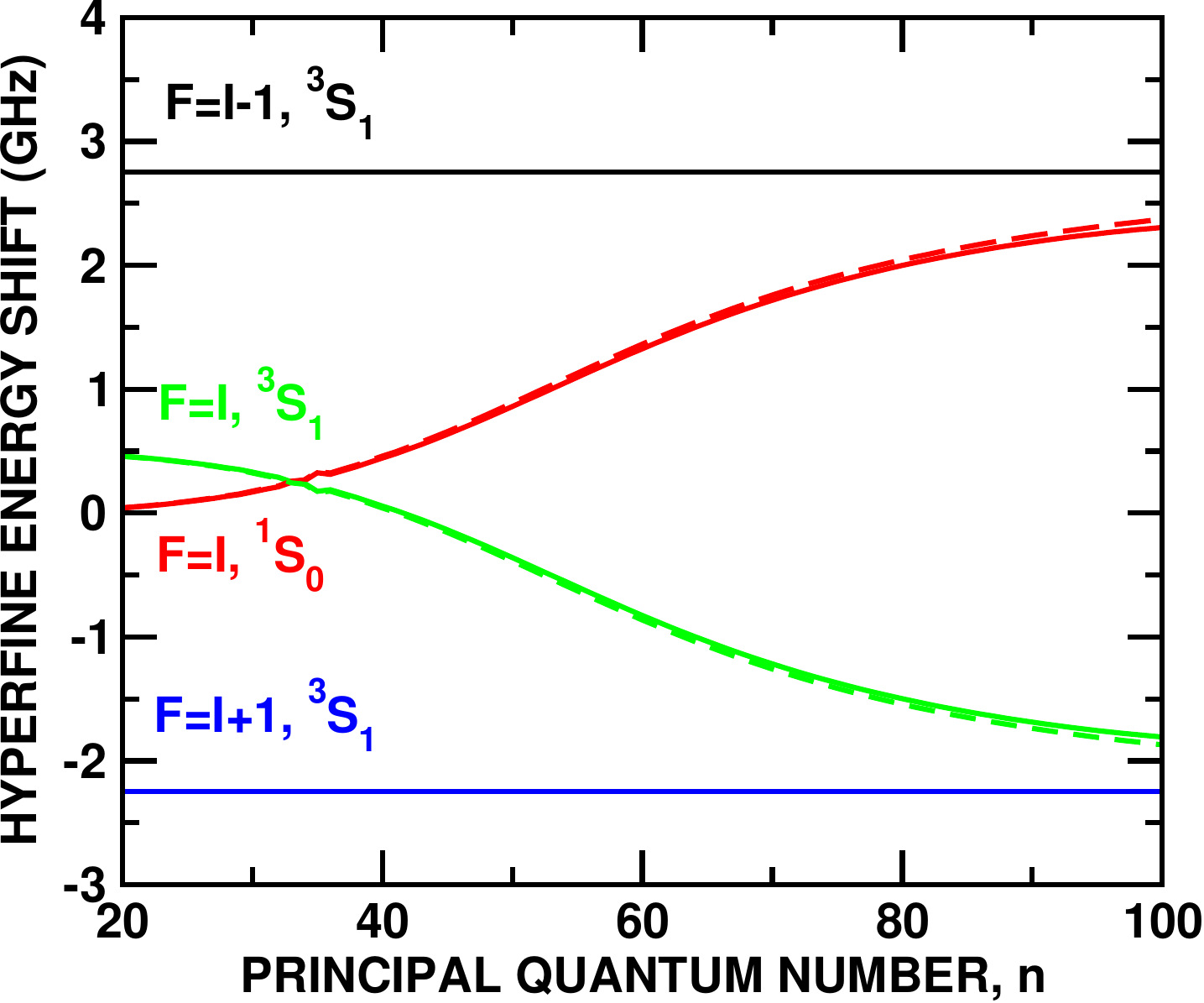}
\end{center}
\caption{\label{fig:shiftS}(Color online.) (Solid lines) Hyperfine energy shifts of the $\nSLJ{5sns}{1,3}{S}{}$ states in \Sr{87} relative to the eigenvalues of $H_0(88,m_{87})$ [see Eq.~(\ref{eq:hamil})]. The state labels for the mixed $F=I$ sub-manifold [{Eqs.~(\ref{eq:hfs1}-\ref{eq:hfs3})}] indicate the state with the largest overlap. (Dashed lines) Hyperfine energy shifts for ${F=I}$ states when mixing of adjacent $n$ levels due to the hyperfine interaction is neglected, i.e., setting ${O_{n,n'} = \delta_{n,n'}}$ in Eq.~(\ref{eq:hfs3}).}
\end{figure}

Using this hyperfine interaction matrix together with the Hamiltonian $H_0(88,m_{87})$ derived from the measured energies for ${n \le 70}$ $\SLJ{1}{S}{0}$ states \cite{beig82} and for ${n \le 40}$ $\SLJ{3}{S}{1}$ states \cite{beig82b} in \Sr{88} as well as values obtained by extrapolation \cite{vail12} to higher $n$ using the Rydberg-Ritz formula, the Hamiltonian [Eq.~(\ref{eq:hamil})] is diagonalized. 
(Note that the Rydberg-Ritz formula is also used for {low-$n$} states when the measured data show large fluctuations.) 
$H_0(88,m_{87})$ is constructed by first converting the measured energies and ionization threshold \cite{sans10} for \Sr{88} to quantum defects using Eq.~(\ref{eq:ene0}) with the Rydberg constant $R(m_{88})$ mass-scaled for \Sr{88}.
These quantum defects are then converted back to energies appropriate to \Sr{87} using the ionization threshold for $\Sr{87}$ and the corresponding \Sr{87} mass-scaled Rydberg constant $R(m_{87})$. 
The ionization threshold for \Sr{87} has only been measured for the $\nSLJF{5s}{2}{S}{1/2}{4}$ state. 
The threshold $E^{(0)}_{\rm ion}$ is therefore estimated by subtracting the hyperfine shift ${-(1/2) a_{5s} (I+1)}$ [{Eqs.~(\ref{eq:eion}, \ref{eq:hfs-1})}] from the measured value. 
Figure~\ref{fig:shiftS} shows the calculated hyperfine shift ${E - E_{n,S,L,J}^{(0)}}$ where $E$ is an eigenenergy of the Hamiltonian $H(87)$. 
As reference we use the eigenvalues $E_{n,S,L,J}^{(0)}$ of $H_0(88,m_{87})$. 
In the case of singlet-triplet mixing (for ${F=I}$) we use the eigenvalue of the $\SLJ{}{S}{}$ state that features the largest overlap. 
For {low-$n$} states, the hyperfine interaction is much smaller than the singlet-triplet splitting. 
Therefore, the hyperfine interaction can be treated perturbatively and the first-order term in the energy shift vanishes for $\SLJ{1}{S}{0}$ states [Eq.~(\ref{eq:hfs1})] and is ${-(1/2) a_{5s} \simeq \SI{0.5}{\GHz}}$ for $\SLJ{3}{S}{1}$ states [Eq.~(\ref{eq:hfs2})] as observed in Fig.~\ref{fig:shiftS} for ${n \simeq 20}$.
As $n$ increases, the mixing of the singlet and triplet states leads to strong deviations from the perturbative estimates and eventually, in the high $n$ limit, the shifts of the two ${F=I}$ states approach either that of the ${F=I+1}$ or of the ${F=I-1}$ state, the splitting of which corresponds to that of the ionization limits. 
For very high $n$ the inter-$n$ mixing becomes non-negligible. 
The comparison between the full calculation and the one in which inter-$n$ mixing is switched off (i.e., ${O_{n,n'} = \delta_{n,n'}}$ in Eq.~(\ref{eq:hfs3})), also shown in Fig.~\ref{fig:shiftS}, reveals that only for ${n > 80}$ do the contributions from different $n$ levels become visible. 
Around ${n=100}$, the difference between the two calculations is \SI{\sim 70}{\MHz}. 
We note that the accuracy of the calculations is limited by the uncertainties in the measurement of the Rydberg states and the ionization thresholds as well as by the Rydberg-Ritz fitting used to derive the energies $E^{(0)}_{n,S,L,J}$. 
An order of magnitude estimate of the uncertainty can be obtained as follows. 
Taking, for example, the measured data \cite{beig82b} for ${n \le 40}$ with an accuracy of ${\SI{0.01}{\per\cm} \simeq \SI{300}{\MHz}}$, this uncertainty translates into an error of, at most, \num{0.002} in the quantum defect. 
For high $n$, assuming that the quantum defect can be extrapolated with the same accuracy of \num{0.002}, the resulting error in high Rydberg states would be $0.002/n^3$ corresponding to \SI{\sim 35}{\MHz} for ${n \sim 70}$ and \SI{\sim 13}{\MHz} for ${n \sim 100}$.
 
\subsection{Energy shift of $\SLJ{}{D}{}$ states}

Extending the method used for the $\SLJ{}{S}{}$ states to $\SLJ{}{D}{}$ states presents considerable difficulties. 
The available measured levels for the \SLJ{3}{D}{} states of \Sr{88} are limited to ${n \lesssim 40}$ \cite{beig82b}. 
Moreover, the quantum defects extracted from these measurements feature a non-negligible $n$ dependence which precludes the accurate extrapolation to very {high-$n$} states. 
In fact, attempts to employ quantum defects derived from earlier measurements of {low-$n$} states \cite{vail12} to describe the present data for higher $n$ failed to provide any reasonable degree of agreement. 
Therefore, for the $\SLJ{3}{D}{}$ states we apply the method outlined above, only in reverse. 
Following Eq.~(\ref{eq:hamil}), we use the present experimental data for \Sr{87} to determine spectroscopic information for the bosonic isotope. 
In practice, the quantum defects $\mu^{(0)}_{n,S,L,J}$ [Eq.~(\ref{eq:ene0})] are treated initially as free parameters and the eigenvalues of $H(87)$ are evaluated for each guess of $\mu^{(0)}_{n,S,L,J}$. 
By scanning through the parameter space in $\mu^{(0)}_{n,S,L,J}$ the set of quantum defects that yield, for the hyperfine energy levels of \Sr{87}, the best agreement with the measured data are identified. 
The quantum defects for the {$n=50, 60$ and $98$} levels obtained in this manner are used to update the Rydberg-Ritz formula for the $\SLJ{3}{D}{}$ states, in particular for their high $n$ limits. 
These quantum defects are then tested against data for {${n \simeq 50}$ and $80$} Rydberg states in \Sr{88}. 
Moreover, the updated Rydberg-Ritz formula can be used to calculate the hyperfine structure for {higher-$n$} \Sr{87} Rydberg $\SLJ{}{D}{}$ states and the resulting predictions tested against measured data for {high-$n$} (${n \sim 100, 280}$) $\SLJ{}{D}{}$ states \cite{sun89,ye13}. 
In our analysis, we include all singlet and triplet $\SLJ{}{D}{}$ states, i.e., $\ket{((5snd) \, \SLJ{1}{D}{2}, I) F}$ and $\ket{((5snd) \, \SLJ{3}{D}{1,2,3}, I) F}$ states with ${|I - J| \le F \le I+J}$. 

For Rydberg $\SLJ{}{D}{}$ states, the spin-orbit interaction (see Fig.~\ref{fig:deltaE}) leads to a breakdown of the $LS$ coupling even in the absence of nuclear spin. 
This small but non-negligible coupling induces a weak mixing between the $\SLJ{1}{D}{2}$ and the $\SLJ{3}{D}{2}$ states \cite{eshe77, sun89}. 
To account for this mixing, the $\SLJ{}{D}{}$ states for ${I=0}$, i.e., eigenstates of the Hamiltonian $H_0(88,m_{87})$, are expanded as
\begin{eqnarray}
\ket{(5snd) \, \SLJ{1}{D}{2}} &=& \cos\theta \ket{n_1^* \, \SLJ{1}{D}{2}} + \sin\theta \ket{n_1^* \, \SLJ{3}{D}{2}} \nonumber \\
\ket{(5snd) \, \SLJ{3}{D}{2}} &=& -\sin\theta \ket{n_3^* \, \SLJ{1}{D}{2}} + \cos\theta \ket{n_3^* \, \SLJ{3}{D}{2}} \, .
\label{eq:admix}
\end{eqnarray}
The $\ket{n_{1,3}^* \, \SLJ{1,3}{D}{2}}$ states denote pure singlet and triplet states while the mixed singlet or triplet states are denoted by $\ket{(5snd) \, \SLJ{${2S+1}$}{D}{2}}$. 
With the help of an independent TAE calculation we have verified that the radial wave functions of both pure singlet and triplet states $\ket{n_{2S+1}^* \, \SLJ{1}{D}{2}}$ and $\ket{n_{2S+1}^* \, \SLJ{3}{D}{2}}$ follow the same asymptotic behavior characterized by the same scattering phase shift, or equivalently, effective quantum number ${n^*_{2S+1} = n - \mu^{(0)}_{n,S,L=2,J=2}}$. 
The mixing of singlet and triplet states is known to be strong around ${n=15}$ and the value of $\theta$ is sensitive to the value of $n$ \cite{eshe77}. 
Indeed, the singlet and the triplet states include a sizable admixture of the $4d6s$ configuration around ${n=15}$ modifying the magnitude of the electron-electron interaction. 
Consequently, the spin-orbit interaction becomes comparable to the electron-electron interaction leading to strong mixing of the singlet and triplet states. 
This results in a pronounced deviation of the singlet-triplet splitting from the $n^{-3}$ scaling around ${n=15}$ (Fig.~\ref{fig:deltaE}). 
For higher $n$, on the other hand, the singlet-triplet mixing becomes nearly $n$-independent and $\theta$ is estimated to converge towards ${\theta \sim -0.14}$. 
(The TAE calculation yields a similar value, ${\theta \sim -0.16}$.) 
As will be shown later, the current experimental data can be well reproduced when $\theta$ is set to \num{-0.14} and this value is used in the following calculations. 
Including this admixture, the matrix elements of the hyperfine operator $V_{\rm HF}$ in the $\SLJ{}{D}{}$ sector can be calculated (see Appendix~\ref{app:d-state-eqs}). 

\begin{figure}[!htbp]
\begin{center}
\includegraphics[width=8.6cm, keepaspectratio=true]{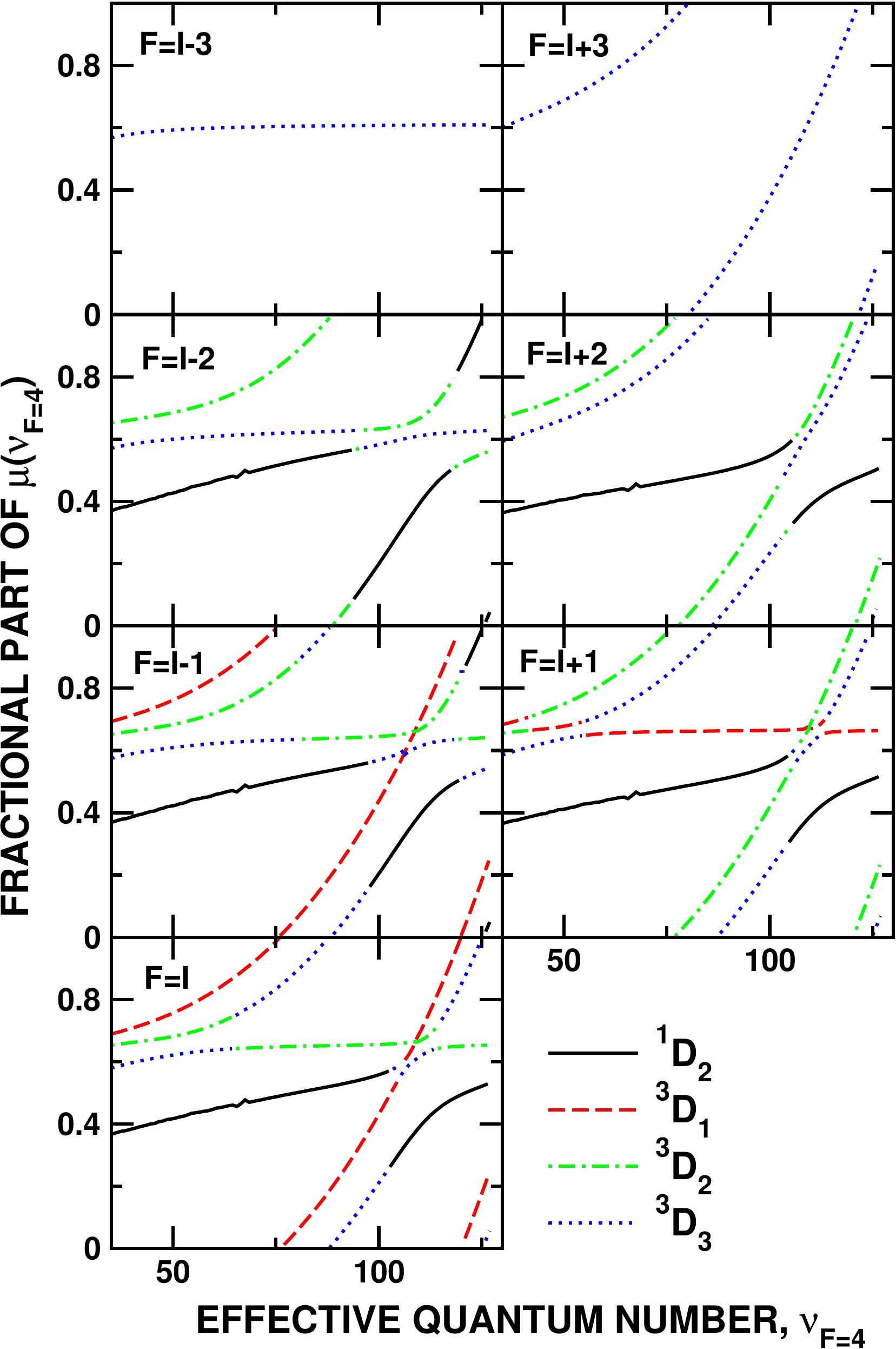}
\end{center}
\caption{\label{fig:shiftD}(Color online.) (Solid lines) Fractional part of quantum defect $\mu(\nu_{F=4})$ evaluated relative to the ${F=4}$ ionization threshold [see Eq.~(\ref{eq:qd2})] as a function of the effective quantum number $\nu_{F=4}$ for different $F$ manifolds of Sr in the $\SLJ{}{D}{}$ sector. Each state is labeled by its dominant $\SLJ{${2S+1}$}{D}{$J$}$ state component.}
\end{figure}

Using the measured quantum defects for \Sr{88} \cite{beig82, beig82b} and the Rydberg-Ritz formula, the hyperfine structure is calculated and plotted in terms of quantum defects (see Fig.~\ref{fig:shiftD}). 
This quantum defect should converge to a constant value as ${n \to \infty}$ provided that the Rydberg series is pure, i.e., converges to a well-defined ionization threshold. 
However, since for strontium two ionization limits ${E_{\rm ion}(F=4 \mbox{ and } 5)}$ [Eq.~(\ref{eq:eion})] are present and the channels are strongly mixed by the hyperfine interaction, it is not straightforward to identify the proper ionization limit for each Rydberg series. 
We illustrate this point in Fig.~\ref{fig:shiftD} where the fractional part of the quantum defect (${\mu \mod 1}$) relative to just one of the two thresholds, ${E_{\rm ion}(F=4)}$, is plotted. 
The quantum defect relative to ${E_{\rm ion}(F=4)}$ is defined as
\begin{equation}
\mu(\nu_{F=4}) = n - \nu_{F=4} \quad
\mbox{ with } \quad
\nu_{F=4} = \sqrt{\frac{R(m_{87})}{E_{\rm ion}(F=4)-E}} \, ,
\label{eq:qd2}
\end{equation}
where $E$ is the eigenenergy of the Hamiltonian $H(87)$ [Eq.~(\ref{eq:hamil})] and is expressed in terms of the effective quantum number $\nu_{F=4}$ for the different $F$ manifolds. 
A few different $\nu_{F=4}$ dependences in $\mu(\nu_{F=4})$ can be distinguished: a near constant $\mu(\nu_{F=4})$ as seen for ${F=I-3}$ indicates convergence to ${E_{\rm ion}(F=4)}$, and a monotonically increasing $\mu(\nu_{F=4})$ (${F = I+3}$) signals the approach of the other ionization threshold ${E_{\rm ion}(F=5)}$,
\begin{eqnarray}
\mu(\nu_{F=4})
&=& n -  \sqrt{\frac{R_{87}}{E_{\rm ion}(F=5) + \Delta E_{\rm ion} -E}} \nonumber \\
&\simeq& \mu(\nu_{F=5}) + \frac{\Delta E_{\rm ion}}{2 R_{87}} \nu_{F=5}^3
\label{eq:qd1}
\end{eqnarray}
with ${\nu_{F=5} = [R_{87}/(E_{\rm ion}(F=5)-E)]^{1/2}}$ and ${\Delta E_{\rm ion} = E_{\rm ion}(F=4) - E_{\rm ion}(F=5) > 0}$. 
In the {high-$n$} limit, while $\mu(\nu_{F=5})$ becomes a constant, $\mu(\nu_{F=4})$ increases with $n$. 
Around ${\nu_{F=4} \simeq 110}$, $\Delta E_{\rm ion}$ becomes comparable to $n^{-3}$ and the quantum defect will be shifted by $1$ (equivalent to approaching the same value for its fractional part) compared to its value for lower $n$. 
Consequently, the inter-$n$ mixing becomes strong and, correspondingly, the formation of avoided crossings is clearly observed. 
The existence of multiple thresholds affects the extraction of proper quantum defects as for high $n$ the hyperfine interaction can become comparable to the energy splittings between states with ${\Delta n \simeq 1}$ and the asymptotic behavior of the quantum defects may become even more complicated.

\section{Experimental method}

\begin{figure}[!htbp]
\includegraphics[width=8.6cm, keepaspectratio=true]{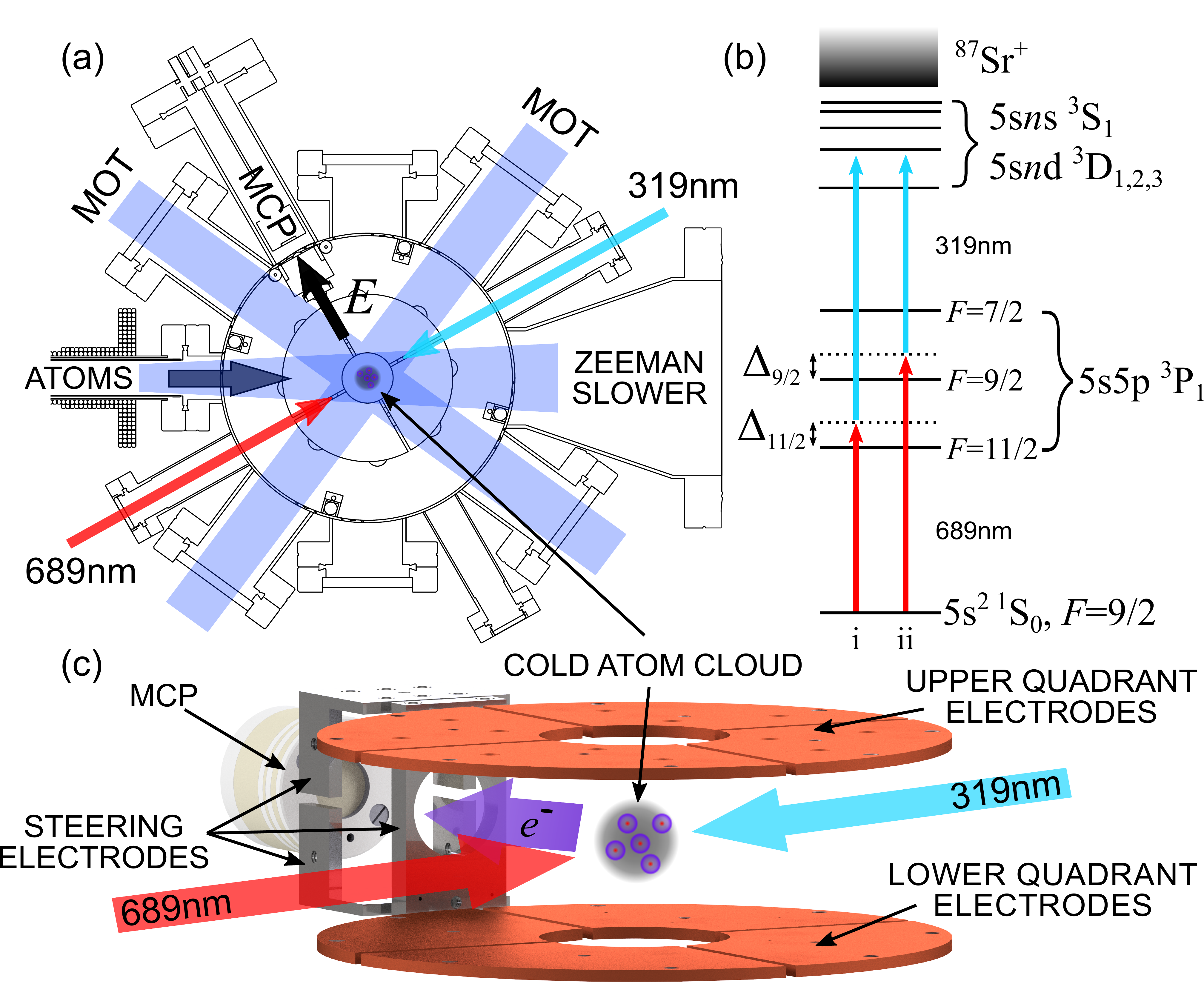}
\caption{\label{fig:exp_setup}(Color online.) (a) Diagram of the experimental arrangement showing the \SI{461}{\nm} cooling beams and the counter-propagating \SI{689}{\nm} and \SI{319}{\nm} Rydberg excitation lasers. (b) Two-photon excitation scheme utilizing either the {(\romannumeral 1)} $\nSLJF{5s5p}{3}{P}{1}{11/2}$ or {(\romannumeral 2)} $\nSLJF{5s5p}{3}{P}{1}{9/2}$ intermediate states. The detunings ${\Delta_{11/2} \sim \SI{12}{\MHz}}$ and ${\Delta_{9/2} \sim \SI{36}{\MHz}}$ remain fixed. (c) Arrangement of the electrodes used for ionizing Rydberg atoms and guiding the electrons towards the MCP detector.}
\end{figure}

A schematic diagram of the present experimental arrangement is presented in Fig.~\ref{fig:exp_setup}. 
The cooling and trapping of strontium is described in detail elsewhere \cite{Xu_2003, Nagel_2003, Mukaiyama_2003, DeSalvo_2010, Stellmer_2013}. 
Briefly, starting from a Zeeman slowed atomic beam, \Sr{87} atoms are first cooled and trapped using a ``blue'' magneto-optical trap (MOT) operating on the \SI{461}{\nm} ${\nSLJ{5s^2}{1}{S}{0} \rightarrow \nSLJ{5s5p}{1}{P}{1}}$ transition. 
The atoms are then further cooled in a narrow-line ``red'' MOT utilizing the ${\nSLJ{5s^2}{1}{S}{0} \rightarrow \nSLJ{5s5p}{3}{P}{1}}$ intercombination line at \SI{689}{\nm}. 
Approximately \SI{e6}{\atoms} at \SI{\sim2}{\micro\K} are captured before turning off all trapping fields for spectroscopy measurements. 

Rydberg atoms are created by two-photon excitation using counter-propagating cross-linearly-polarized \SI{689}{\nm} and \SI{319}{\nm} laser beams which drive transitions to the $\nSLJ{5sns}{3}{S}{1}$ and $\nSLJ{5snd}{3}{D}{1,2,3}$ Rydberg levels via the intermediate {$\nSLJF{5s5p}{3}{P}{1}{9/2}$ or $11/2$} states. 
These intermediate states were selected to take advantage of selection rules to aid in identifying the Rydberg hyperfine states populated (see Fig.~\ref{fig:exp_setup}b). 
The typical detunings of the \SI{689}{\nm} laser were ${\Delta_{9/2} \sim \SI{36}{\MHz}}$ and ${\Delta_{11/2} \sim \SI{12}{\MHz}}$. 
The \SI{689}{\nm} laser was chopped into \SIrange{10}{20}{\us}-long pulses to generate temporally-localized groups of Rydberg atoms. 
The number of Rydberg atoms produced by each pulse was determined by using the electrodes in Fig.~\ref{fig:exp_setup}c to generate a ramped electric field sufficient to ionize the Rydberg atoms. 
The resulting electrons were directed towards, and detected by, a microchannel plate (MCP) whose output was fed into a multichannel scalar (MCS). 
Typically \SIrange{100}{500}{excitation/detection} cycles were performed before loading a new sample and changing the \SI{319}{\nm} laser frequency. 
The stray fields in the trapping region were determined to be less than \SI{10}{\mV\per\cm} and any resultant Stark shifts should therefore be at most a few \si{\MHz} even at ${n \sim 90}$.

The \SI{319}{\nm} radiation was generated by frequency doubling the output of a \SI{638}{\nm} optical parametric oscillator (OPO). 
A sample of the output is sent though a broadband fiber electro-optic modulator (fEOM) from which one of the sidebands was locked to a transfer cavity, allowing the \SI{319}{\nm} laser to be scanned over multiple \si{\GHz}. 
The transfer cavity was stabilized using a \SI{689}{\nm} master laser locked to the ${\nSLJ{5s^2}{1}{S}{0} \rightarrow \nSLJ{5s5p}{3}{P}{1}}$ transition in \Sr{88}. 
The linewidth of the \SI{319}{\nm} laser is estimated to be \SI[input-comparators=\lesssim]{\lesssim500}{\kHz} based on the narrowest observed spectroscopic features.

\begin{figure}[!htbp]
\includegraphics[width=8.6cm, keepaspectratio=true]{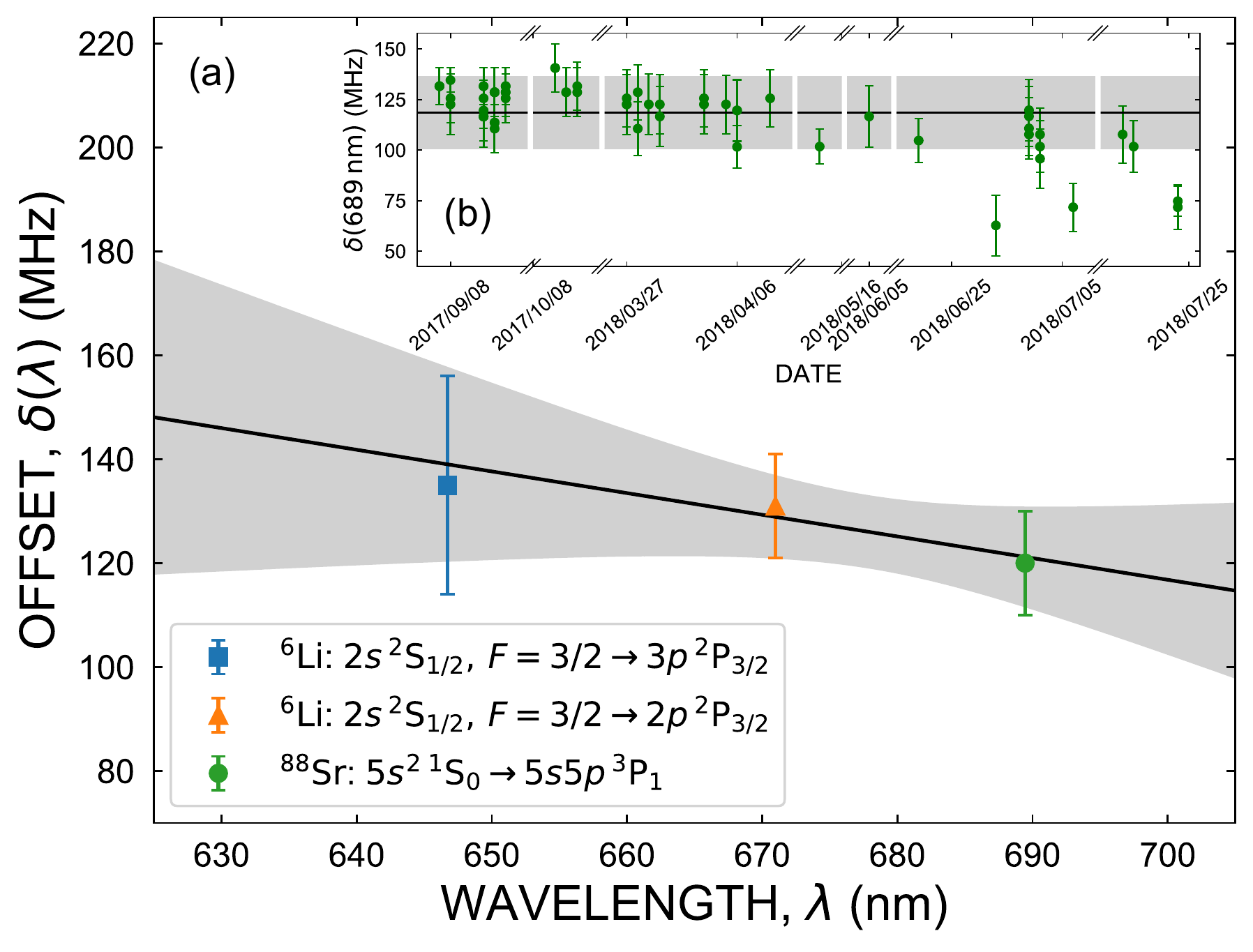}
\caption{\label{fig:wm_cal}(Color online.) (a) Wavelength dependence of the offset ($\delta$) between the measured and published transition frequencies used to calibrate the wavemeter: (black line) linear fit used to obtain the offset at \SI{638}{\nm}, (shaded region) uncertainty in the wavemeter calibration obtained from Monte Carlo simulations (see text). (b, inset) Offset of the \SI{689}{\nm} transition in \Sr{88} measured at different times.}
\end{figure}

A wavemeter (EXFO WA-1500) was used to measure the wavelength of the \SI{638}{\nm} output from the OPO and hence determine the Rydberg state energies with a resolution-limited statistical uncertainty ($\sigma_\mathrm{stat}$) of about \SI{\pm 15}{\MHz} ($\SI{\pm 30}{\MHz}$) at \SI{638}{\nm} (\SI{319}{\nm}). 
In order to estimate systematic offsets in the wavemeter, the frequencies of lasers locked to atomic transitions in \Sr{88} (${\nSLJ{5s^2}{1}{S}{0} \rightarrow \nSLJ{5s5p}{3}{P}{1}}$ at \SI{689}{\nm} \cite{fcd03, sans10}) and in \Li{6} (${\nSLJF{2s}{2}{S}{1/2}{3/2} \rightarrow \nSLJ{2p}{2}{P}{3/2}}$ at \SI{671}{\nm} and ${\nSLJF{2s}{2}{S}{1/2}{3/2} \rightarrow \nSLJ{3p}{2}{P}{3/2}}$ at $\SI{646}{\nm}/2=\SI{323}{\nm}$ \cite{Sansonetti_2011, Sansonetti_2012, reb95}). 
The measured wavelengths were then compared to the published values for the same transitions and the differences, $\delta$, between the measured and published frequencies are shown in Fig.~\ref{fig:wm_cal}. 
A linear fit yields a correction of \SI{\approx 140}{\MHz} at \SI{638}{\nm}. 
In an attempt to estimate the systematic uncertainty in this calibration factor, a Monte Carlo sampling was adopted in which linear fits to points drawn at random from the Gaussian uncertainty distributions appropriate to each point in the calibration were repeated, resulting in a systematic uncertainty ($\sigma_\mathrm{sys}$) of about \SI{\pm 25}{\MHz} ($\SI{\pm 50}{\MHz}$) at \SI{638}{\nm} (\SI{319}{\nm}). 
To check for drifts in the wavemeter calibration, each \SI{638}{\nm} wavelength measurement was followed by a reference measurement of the \SI{689}{\nm} master laser. 
As shown in the inset Fig.~\ref{fig:wm_cal}b, the day-to-day variations were relatively small compared to the wavemeter's systematic uncertainty. 
Whereas our wavemeter limits the measurements of individual term energies to \SI{\sim 60}{\MHz}, line separations can be measured to \si{\kHz}-level accuracies when scanning within a single free spectral range (FSR) of the transfer cavity, and to \si{\MHz}-level accuracies when piecing together scans over successive FSRs.

\section{Results and discussion}

\begin{figure}[!htbp]
\includegraphics[width=8.6cm, keepaspectratio=true]{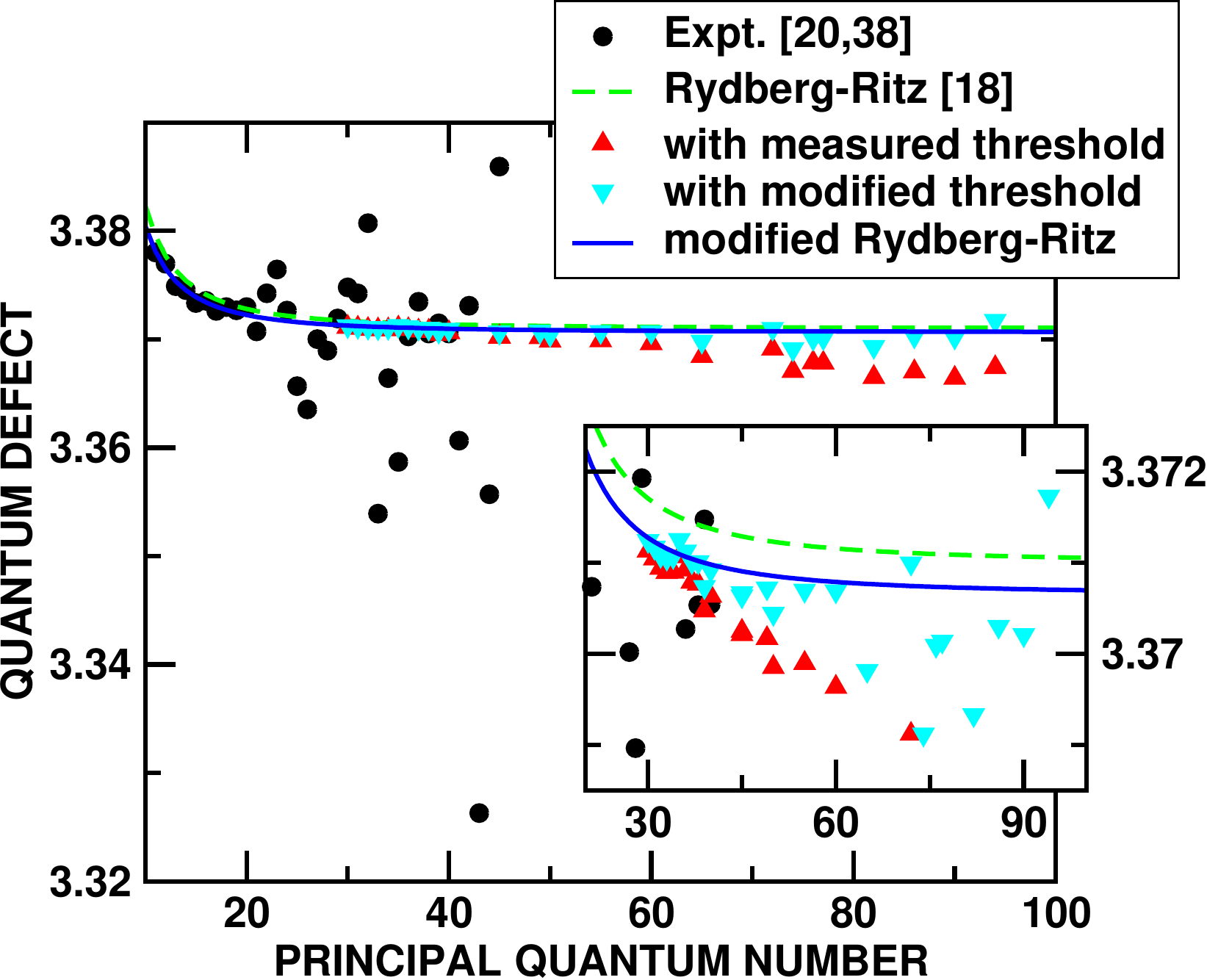}
\caption{\label{fig:Rydberg-Ritz-S} (Color online.) Quantum defects $\mu^{(0)}_{n,S,L,J}$ for the $\nSLJ{5sns}{3}{S}{1}$ levels: ($\bullet$) measurements from earlier work \cite{beig82,beig82b}, ({\color{red} $\blacktriangle$}) present measurements of the $\nSLJF{5sns}{3}{S}{1}{11/2}$ states derived from the earlier ionization limit \cite{sans10}, ({\color[rgb]{0,1,1} $\blacktriangledown$}) present measurements with modified ionization limit (see text). ({\color{green}\dashedline}) predictions using the Rydberg-Ritz formulae from \cite{vail12} and ({\color{blue}\solidline}) the modified Rydberg-Ritz formulae. The inset shows the {higher-$n$} region on an expanded scale.}
\end{figure}

Table~\ref{tab:data-s-state} lists the measured term energies for multiple $\nSLJ{5sns}{1,3}{S}{1}$ states with ${30 \lesssim n \lesssim 99}$. 
Figure~\ref{fig:Rydberg-Ritz-S} shows quantum defects $\mu^{(0)}_{n,S,L,J}$ for the $\nSLJ{5sns}{3}{S}{1}$ states either measured for \Sr{88} \cite{beig82,beig82b} or obtained using the corresponding Rydberg-Ritz formula \cite{vail12} together with those extracted from the current measurement of the $\nSLJF{5sns}{3}{S}{1}{11/2}$ states for \Sr{87}. 
Since the hyperfine energy shift for the $\nSLJF{5sns}{3}{S}{1}{11/2}$ states is constant [Eq.~(\ref{eq:hfs+1})], the quantum defects $\mu^{(0)}_{n,S,L,J}$ of the corresponding bosonic isotope can be uniquely determined. 
The quantum defects obtained in this manner deviate from the values predicted by the earlier Rydberg-Ritz formula displaying a slow decrease in $\mu^{(0)}_{n,S,L,J}$ with increasing $n$. 
In line with the earlier discussion [Eq.~(\ref{eq:qd1})], such a systematic decrease in $\mu^{(0)}_{n,S,L,J}$ with $n$ is typically observed when the ionization threshold is slightly shifted. 
In the current study the previously reported ionization threshold for \Sr{87} \cite{beig82, sans10} is used in Eq.~(\ref{eq:ene0}) to convert between the energy and the quantum defect. 
After subtracting the hyperfine energy correction its value is ${E_{\rm ion}^{(0)}=\SI{45932.1943}{\per\cm}}$. 
The present measured energy levels can be converted, on average, to a converged, nearly constant quantum defect if a slightly higher threshold energy ${E_{\rm ion}^{(0)} \simeq \SI{45932.1956}{\per\cm}}$ is used (see Fig.~\ref{fig:Rydberg-Ritz-S}). 
This would correspond to an energy shift of \SI{\sim 40}{\MHz}. 
(We note that other sources of uncertainty such as specific isotope effects, mass polarization contributions, or stray field effects can be ruled out.)
Due to the fluctuations in the measured quantum defects (Fig.~\ref{fig:Rydberg-Ritz-S}) for {high-$n$}, the ionization threshold can be determined only within an error of \SI{\sim\pm 20}{\MHz}. 

Another feature observed in Fig.~\ref{fig:Rydberg-Ritz-S} is a shift of the measured $\mu^{(0)}_{n,S,L,J}$ from the earlier Rydberg-Ritz prediction. 
In particular, since for low-lying states, ${30 < n < 40}$, the quantum defects are insensitive to small differences in the ionization threshold, this is not true at high $n$ and the observed shift suggests the Rydberg-Ritz formula for the $\SLJ{3}{S}{}$ states needs to be updated. 
The combined data from the earlier measurements \cite{beig82,beig82b} for \Sr{88} and the current measurements for \Sr{87} can be well fit using the Rydberg-Ritz expression
\begin{equation}
\label{eq:Rydberg-Ritz}
\mu^{(0)}_{n,S,L,J}=\mu_{0} + \frac{\alpha}{(n-\mu_{0})^{2}}  + \frac{\beta}{(n-\mu_{0})^{4}} \, .
\end{equation}
and the values of $\mu_{0}$, $\alpha$, and $\beta$ given in Table~\ref{tab:delta}, which also includes the corresponding values derived from the earlier measurements at lower $n$ \cite{vail12}. 
The change in quantum defect is small (\num{\sim 0.0035}) but, when converted to energy, the difference can be non-negligible for low $n$ states (\SI{\sim80}{\MHz} for ${n = 30}$). 
Table~\ref{tab:data-s-state} includes theoretical predictions based on diagonalization of the rescaled Hamiltonian [Eq.~(\ref{eq:hamil})]. 
The calculations use the modified Rydberg-Ritz formula for $\mu^{(0)}_{n,S,L,J}$ together with the measured ionization threshold \cite{beig82, sans10}.
On average, the present theoretical estimates lie slightly below the measured energy levels and, in the {high-$n$} limit, their differences converge to a near-constant value of \SIrange[range-phrase = --]{40}{50}{\MHz}. 
This provides another indication that the ionization threshold should be modified. 

To remove the uncertainty in the ionization limit from the comparison between experiment and theory, we also include in Table~\ref{tab:data-s-state} the measured energy differences between the $\nSLJF{5sns}{1}{S}{0}{9/2}$ or the {$\nSLJF{5sns}{3}{S}{1}{7/2, 9/2}$} states and the corresponding $\nSLJF{5sns}{3}{S}{1}{11/2}$ states together with the values predicted by theory.
As seen in Table~\ref{tab:data-s-state}, the discrepancies between these values are typically well below \SI{0.0005}{\per\cm}, \SI[input-comparators=\simeq]{\simeq 15}{\MHz}. 
Therefore, in the following, we focus on relative energies in our analysis of $\SLJ{}{D}{}$ states.

\begin{table*}[!htbp]
\caption{\label{tab:data-s-state}Experimentally measured and calculated energies of selected $\nSLJ{5sns}{1}{S}{0}$ and $\nSLJ{5sns}{3}{S}{1}$ states in \Sr{87}. $\Delta E_\text{exp}$ and $\Delta E_\text{th}$ are the measured and predicted separations from the $\nSLJF{5sns}{3}{S}{1}{11/2}$ state of the same $n$ which is used as a reference. The uncertainties shown include both the statistical and systematic uncertainties in the wavemeter calibration.}
\begin{ruledtabular}
\begin{tabular}{ccccccccc}
Series	& $n$					& Term				& $F$		& $E_\text{exp}$ (\si{\per\cm})	& $\Delta E_\text{exp}$ (\si{\GHz})	& $E_\text{th}$	(\si{\per\cm})	& $\Delta E_\text{th}$ (\si{\GHz})	\\ \hline
$5sns$	& $40$					& $\SLJ{1}{S}{0}$	& $9/2$		& \num{45850.8762+-0.0021}		& \num{16.35+-0.08}					& \num{45850.8702}				& \num{16.22}						\\
		& $60$					& 					&			& \num{45898.1444+-0.0022}		& \num{7.28+-0.09}					& \num{45898.1421}				& \num{7.26}						\\
		& $72$					& 					&			& \num{45909.0252+-0.0020}		& \num{6.10+-0.09}					& \num{45909.0240}				& \num{6.1}							\\
		& $74$					& 					&			& \num{45910.3230+-0.0021}		& \num{5.98+-0.09}					& \num{45910.3211}				& \num{5.99}						\\
		& $76$					& 					&			& \num{45911.5148+-0.0020}		& \num{5.91+-0.08}					& \num{45911.5127}				& \num{5.89}						\\
		& $77$					& 					&			& \num{45912.0738+-0.0020}		& \num{5.84+-0.09}					& \num{45912.0725}				& \num{5.85}						\\
		& $78$					& 					&			& \num{45912.6114+-0.0020}		&									& \num{45912.6100}				& \num{5.81}						\\
		& $82$					& 					&			& \num{45914.5606+-0.0022}		& \num{5.66+-0.09}					& \num{45914.5589}				& \num{5.67}						\\
		& $86$					& 					&			& \num{45916.2336+-0.0021}		& \num{5.56+-0.08}					& \num{45916.2321}				& \num{5.56}						\\
		& $90$					& 					&			& \num{45917.6802+-0.0019}		& \num{5.46+-0.08}					& \num{45917.6791}				& \num{5.47}						\\
		& $94$					& 					&			& \num{45918.9402+-0.0019}		& \num{5.40+-0.08}					& \num{45918.9388}				& \num{5.39}						\\
		& $98$\footnotemark[1]	& 					&			& \num{45920.0438+-0.0022}		& \num{5.325+-0.005}				& \num{45920.0423}				& \num{5.327}						\\	\hline
$5sns$	& $40$					& $\SLJ{3}{S}{1}$	& $7/2$		& \num{45850.4974+-0.0021}		& \num{4.99+-0.08}					& \num{45850.4960}				& \num{5.0}							\\
		& $60$					& 					&			& \num{45898.0688+-0.0021}		& \num{5.02+-0.08}					& \num{45898.0668}				& \num{5.0}							\\	\hline
$5sns$	& $40$					& $\SLJ{3}{S}{1}$	& $9/2$		& \num{45850.4078+-0.0021}		& \num{2.31+-0.08}					& \num{45850.4061}				& \num{2.31}						\\
		& $50$					& 					&			& \num{45881.7138+-0.0022}		& \num{1.88+-0.09}					& \num{45881.7119}				& \num{1.89}						\\
		& $72$					& 					&			& \num{45908.8546+-0.0021}		& \num{0.99+-0.09}					& \num{45908.8528}				& \num{0.97}						\\
		& $74$					& 					&			& \num{45910.1518+-0.0022}		& \num{0.85+-0.09}					& \num{45910.1516}				& \num{0.91}						\\
		& $76$					& 					&			& \num{45911.3460+-0.0019}		& \num{0.85+-0.08}					& \num{45911.3445}				& \num{0.85}						\\
		& $77$					& 					&			& \num{45911.9068+-0.0021}		& \num{0.83+-0.09}					& \num{45911.9049}				& \num{0.83}						\\
		& $78$					& 					&			& \num{45912.4444+-0.0019}		&									& \num{45912.4429}				& \num{0.8}							\\
		& $82$					& 					&			& \num{45914.3958+-0.0021}		& \num{0.72+-0.09}					& \num{45914.3935}				& \num{0.71}						\\
		& $86$					& 					&			& \num{45916.0696+-0.0021}		& \num{0.64+-0.08}					& \num{45916.0677}				& \num{0.63}						\\
		& $90$					& 					&			& \num{45917.5172+-0.0021}		& \num{0.57+-0.08}					& \num{45917.5155}				& \num{0.56}						\\
		& $94$					& 					&			& \num{45918.7774+-0.0022}		& \num{0.52+-0.09}					& \num{45918.7759}				& \num{0.51}						\\
		& $98$\footnotemark[1]	& 					&			& \num{45919.8816+-0.0022}		& \num{0.46302+-0.00007}			& \num{45919.8800}				& \num{0.46164}						\\	\hline
$5sns$	& $30$					& $\SLJ{3}{S}{1}$	& $11/2$	& \num{45777.3637+-0.0020}		&									& \num{45777.3621}				& 									\\
		& $31$					& 					&			& \num{45788.3644+-0.0021}		&									& \num{45788.3624}				& 									\\
		& $32$					& 					&			& \num{45798.2325+-0.0022}		&									& \num{45798.2302}				& 									\\
		& $33$					& 					&			& \num{45807.1179+-0.0019}		&									& \num{45807.1158}				& 									\\
		& $34$					& 					&			& \num{45815.1469+-0.0021}		&									& \num{45815.1452}				& 									\\
		& $35$					& 					&			& \num{45822.4253+-0.0021}		&									& \num{45822.4252}				& 									\\
		& $36$					& 					&			& \num{45829.0469+-0.0020}		&									& \num{45829.0460}				& 									\\
		& $37$					& 					&			& \num{45835.0865+-0.0021}		&									& \num{45835.0851}				& 									\\
		& $38$					& 					&			& \num{45840.6098+-0.0014}		&									& \num{45840.6085}				& 									\\
		& $39$					& 					&			& \num{45845.6759+-0.0022}		&									& \num{45845.6734}				& 									\\
		& $40$					& 					&			& \num{45850.3308+-0.0015}		&									& \num{45850.3291}				& 									\\
		& $42$					& 					&			& \num{45858.5807+-0.0021}		&									& \num{45858.5793}				& 									\\
		& $43$					& 					&			& \num{45862.2455+-0.0020}		&									& \num{45862.2439}				& 									\\
		& $44$					& 					&			& \num{45865.6435+-0.0021}		&									& \num{45865.6413}				& 									\\
		& $45$					& 					&			& \num{45868.7988+-0.0015}		&									& \num{45868.7968}				& 									\\
		& $49$					& 					&			& \num{45879.4140+-0.0019}		&									& \num{45879.4124}				& 									\\
		& $50$					& 					&			& \num{45881.6510+-0.0021}		&									& \num{45881.6488}				& 									\\
		& $55$					& 					&			& \num{45890.9526+-0.0020}		&									& \num{45890.9511}				& 									\\
		& $60$					& 					&			& \num{45897.9014+-0.0019}		&									& \num{45897.9000}					& 									\\
		& $65$					& 					&			& \num{45903.2294+-0.0019}		&									& \num{45903.2272}				& 									\\
		& $72$					& 					&			& \num{45908.8216+-0.0022}		&									& \num{45908.8205}				& 									\\
		& $74$					& 					&			& \num{45910.1236+-0.0022}		&									& \num{45910.1213}				& 									\\
		& $76$					& 					&			& \num{45911.3178+-0.0019}		&									& \num{45911.3161}				& 									\\
		& $77$					& 					&			& \num{45911.8790+-0.0021}		&									& \num{45911.8774}				& 									\\
		& $82$					& 					&			& \num{45914.3718+-0.0022}		&									& \num{45914.3699}				& 									\\
		& $86$					& 					&			& \num{45916.0482+-0.0019}		&									& \num{45916.0467}				& 									\\
		& $90$					& 					&			& \num{45917.4982+-0.0019}		&									& \num{45917.4967}				& 									\\
		& $94$					& 					&			& \num{45918.7600+-0.0021}		&									& \num{45918.7590}				& 									\\
		& $98$					& 					&			& \num{45919.8662+-0.0022}		&									& \num{45919.8646}				& 									\\
		& $99$\footnotemark[1]	& 					&			& \num{45920.1210+-0.0022}		&									& \num{45920.1196}				& 									
\end{tabular}
\end{ruledtabular}
\footnotetext[1]{Measured relative to the $\nSLJF{5s98s}{3}{S}{1}{11/2}$ state, see Table~\ref{tab:data-d-state}.}
\end{table*}

\begin{table*}[!htbp]
\centering
\caption{\label{tab:data-d-state}Comparison of measured and calculated positions of $\nSLJ{5snd}{3}{D}{1,2,3}$ lines for ${n=\text{\numlist{50;60;\sim98}}}$. The splittings $\Delta E_\text{exp}$ between those lines that could be measured during a single FSR scan of the \SI{319}{\nm} laser frequency (delineated by the horizontal lines) or, for ${n \sim 98}$, where neighboring scans could be accurately patched together are included together with the corresponding theoretical predictions. For the ${n=\text{\numrange[range-phrase = --]{98}{99}}}$ scan, all differences are referenced to the $\nSLJF{5sns}{3}{S}{1}{11/2}$ level.}
\begin{ruledtabular}
\begin{tabular}{cccccccc}
Series				& $n$	& Term				& $F$		& $E_\text{exp}$ (\si{\per\cm})	& $\Delta E_\text{exp}$ (\si{\MHz})	& $E_\text{th}$ (\si{\per\cm})	& $\Delta E_\text{th}$ (\si{\MHz})	\\ \hline
$\nSLJ{5snd}{}{}{}$	& $50$	& $\SLJ{3}{D}{1}$	& $7/2$		& \num{45883.1440+-0.0022}		& \num{-295.60+-0.07}				& \num{45883.1414}				& \num{-299.01}						\\
					& $50$	& $\SLJ{3}{D}{1}$	& $9/2$		& \num{45883.1538+-0.0022}		&									& \num{45883.1514}				&									\\
					& $50$	& $\SLJ{3}{D}{2}$	& $11/2$	& \num{45883.1685+-0.0022}		& \num{439.39+-0.07}				& \num{45883.1662}				& \num{443.71}						\\ \hline
$\nSLJ{5snd}{}{}{}$	& $50$	& $\SLJ{3}{D}{2}$	& $7/2$		& \num{45883.2882+-0.0021}		&									& \num{45883.2855}				&									\\
					& $50$	& $\SLJ{3}{D}{2}$	& $9/2$		& \num{45883.2922+-0.0021}		& \num{118.91+-0.07}				& \num{45883.2893}				& \num{114.7}						\\
					& $50$	& $\SLJ{3}{D}{1}$	& $11/2$	& \num{45883.2972+-0.0021}		& \num{269.12+-0.07}				& \num{45883.2942}				& \num{260.55}						\\ \hline
$\nSLJ{5snd}{}{}{}$	& $50$	& $\SLJ{3}{D}{3}$	& $11/2$	& \num{45883.3849+-0.0022}		& \num{-890.64+-0.07}				& \num{45883.3814}				& \num{-890.22}						\\
					& $50$	& $\SLJ{3}{D}{3}$	& $9/2$		& \num{45883.4146+-0.0022}		&									& \num{45883.4111}				&									\\ \hline
$\nSLJ{5snd}{}{}{}$	& $50$	& $\SLJ{3}{D}{3}$	& $7/2$		& \num{45883.4374+-0.0022}		&									& \num{45883.4339}				&									\\ \hline
$\nSLJ{5snd}{}{}{}$	& $60$	& $\SLJ{3}{D}{1}$	& $7/2$		& \num{45898.7367+-0.0021}		& \num{-183.64+-0.07}				& \num{45898.7347}				& \num{-178.89}						\\
					& $60$	& $\SLJ{3}{D}{1}$	& $9/2$		& \num{45898.7428+-0.0021}		&									& \num{45898.7407}				&									\\
					& $60$	& $\SLJ{3}{D}{2}$	& $11/2$	& \num{45898.7521+-0.0021}		& \num{277.34+-0.07}				& \num{45898.7497}				& \num{270.37}						\\ \hline
$\nSLJ{5snd}{}{}{}$	& $60$	& $\SLJ{3}{D}{2}$	& $7/2$		& \num{45898.8568+-0.0022}		& \num{-79.40+-0.07}				& \num{45898.8544}				& \num{-72.67}						\\
					& $60$	& $\SLJ{3}{D}{2}$	& $9/2$		& \num{45898.8594+-0.0022}		&									& \num{45898.8569}				&									\\
					& $60$	& $\SLJ{3}{D}{3}$	& $11/2$	& \num{45898.8618+-0.0022}		& \num{71.37+-0.07}					& \num{45898.8588}				& \num{58.8}						\\ \hline
$\nSLJ{5snd}{}{}{}$	& $60$	& $\SLJ{3}{D}{1}$	& $11/2$	& \num{45898.9223+-0.0022}		& \num{-626.40+-0.07}				& \num{45898.9197}				& \num{-609.77}						\\
					& $60$	& $\SLJ{3}{D}{3}$	& $9/2$		& \num{45898.9432+-0.0022}		&									& \num{45898.9400}				&									\\
					& $60$	& $\SLJ{3}{D}{3}$	& $7/2$		& \num{45898.9608+-0.0022}		& \num{526.18+-0.07}				& \num{45898.9573}				& \num{517.37}						\\ \hline
$\nSLJ{5sns}{}{}{}$	& $98$	& $\SLJ{3}{S}{1}$	& $11/2$	& \num{45919.8662+-0.0022}		&									& \num{45919.8646}				&									\\
					& $98$	& $\SLJ{3}{S}{1}$	& $9/2$		& \num{45919.8816+-0.0022}		& \num{463.02+-0.07}				& \num{45919.8800}				& \num{461.64}						\\
$\nSLJ{5snd}{}{}{}$	& $97$	& $\SLJ{3}{D}{1}$	& $11/2$	& \num{45919.9565+-0.0022}		& \num{2707.6+-3.5}					& \num{45919.9552}				& \num{2716.6}						\\
					& $97$	& $\SLJ{3}{D}{2}$	& $9/2$		& \num{45919.9593+-0.0022}		& \num{2792.4+-3.5}					& \num{45919.9579}				& \num{2796.2}						\\
					& $97$	& $\SLJ{1}{D}{2}$	& $9/2$		& \num{45919.9896+-0.0022}		& \num{3701+-4}						& \num{45919.9879}				& \num{3697}						\\
					& $97$	& $\SLJ{1}{D}{2}$	& $11/2$	& \num{45919.9925+-0.0022}		& \num{3785+-4}						& \num{45919.9909}				& \num{3786}						\\
					& $97$	& $\SLJ{1}{D}{2}$	& $13/2$	& \num{45919.9946+-0.0022}		& \num{3850+-4}						& \num{45919.9933}				& \num{3857}						\\
$\nSLJ{5sns}{}{}{}$	& $98$	& $\SLJ{1}{S}{0}$	& $9/2$		& \num{45920.0438+-0.0022}		& \num{5325+-5}						& \num{45920.0423}				& \num{5327}						\\
$\nSLJ{5snd}{}{}{}$	& $98$	& $\SLJ{3}{D}{1}$	& $9/2$		& \num{45920.0474+-0.0022}		& \num{5432+-5}						& \num{45920.0460}				& \num{5439}						\\
					& $98$	& $\SLJ{3}{D}{2}$	& $11/2$	& \num{45920.0501+-0.0022}		& \num{5512+-5}						& \num{45920.0485}				& \num{5514}						\\
					& $98$	& $\SLJ{3}{D}{2}$	& $13/2$	& \num{45920.0544+-0.0022}		& \num{5641+-5}						& \num{45920.0526}				& \num{5636}						\\
					& $98$	& $\SLJ{3}{D}{3}$	& $13/2$	& \num{45920.0916+-0.0022}		& \num{6756+-6}						& \num{45920.0901}				& \num{6761}						\\
					& $98$	& $\SLJ{3}{D}{3}$	& $11/2$	& \num{45920.0956+-0.0022}		& \num{6877+-6}						& \num{45920.0943}				& \num{6886}						\\
					& $98$	& $\SLJ{3}{D}{3}$	& $9/2$		& \num{45920.0982+-0.0022}		& \num{6954+-6}						& \num{45920.0971}				& \num{6970}						\\
$\nSLJ{5sns}{}{}{}$	& $99$	& $\SLJ{3}{S}{1}$	& $11/2$	& \num{45920.1210+-0.0022}		& \num{7639+-6}						& \num{45920.1196}				& \num{7643}
\end{tabular}
\end{ruledtabular}
\end{table*}

\begin{figure}[!htbp]
\includegraphics[width=8.6cm, keepaspectratio=true]{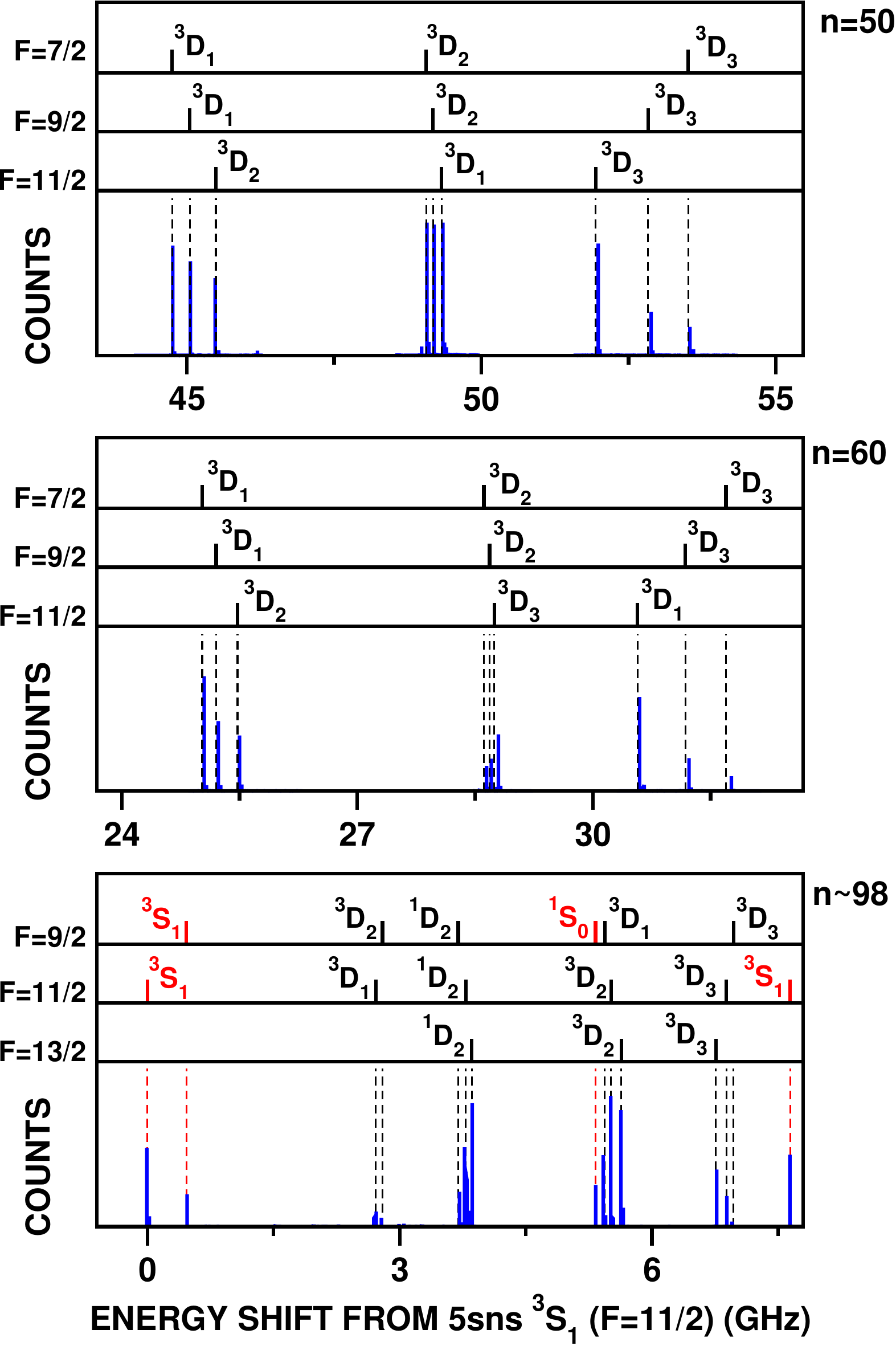}
\caption{\label{fig:spectra_D}(Color online.) (Blue) Measured spectra for $\nSLJ{5snd}{3}{D}{}$ states of \Sr{87} in the vicinity of (a) ${n=50}$, (b) ${n=60}$, (c) ${n=98}$. Energies are given relative to the $5s50s$, $5s60s$, and $5s98s$ $\SLJF{3}{S}{1}{11/2}$ states, respectively. Rydberg excitation was performed following scheme {(\romannumeral 2)} for (a, b) and {(\romannumeral 1)} for (c). The vertical bars above the data show the calculated positions for the various hyperfine states (see text). The measured levels and splittings are given in Table~\ref{tab:data-d-state}.}
\end{figure}

\begin{figure}[!htbp]
\includegraphics[width=8.6cm, keepaspectratio=true]{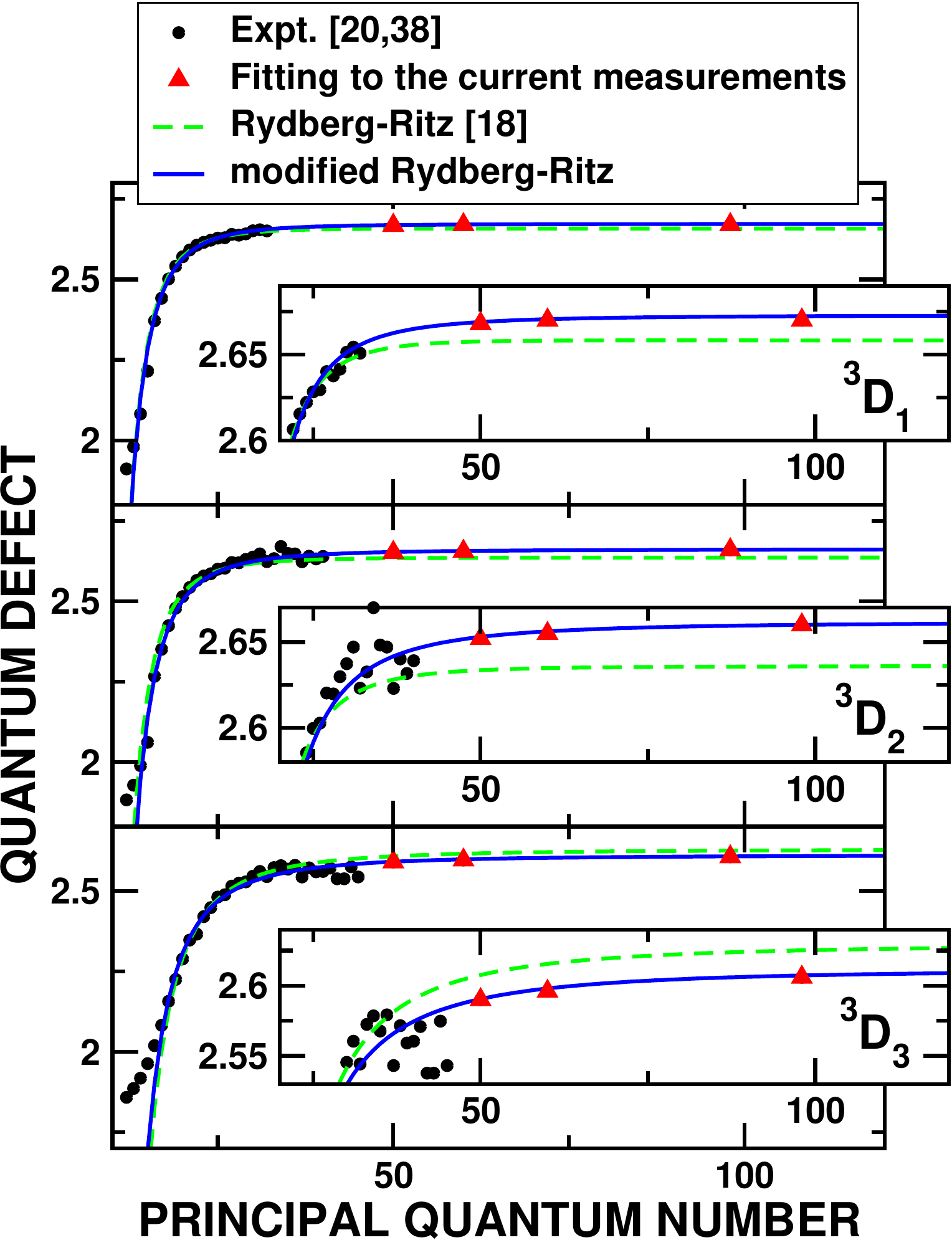}
\caption{\label{fig:Rydberg-Ritz-D}(Color online.) Quantum defects $\mu^{(0)}_{n,S,L,J}$ for the $\nSLJ{5snd}{3}{D}{1,2,3}$ levels: ($\bullet$) measurements from earlier work \cite{beig82,beig82b}; ({\color{red} $\blacktriangle$}) present measurements, ({\color{green}\dashedline}) predictions using the Rydberg-Ritz formulae developed previously \cite{vail12}; ({\color{blue}\solidline}) predictions based on the present updated Rydberg-Ritz formulae (see text). The insets show the {high-$n$} region on an expanded scale.}
\end{figure}

Figure~\ref{fig:spectra_D} shows the positions of the measured $\nSLJ{5snd}{3}{D}{}$ spectral lines for ${n=\text{\numlist{50;60;97;98}}}$ relative to the energy of the $\nSLJF{5sns}{3}{S}{1}{11/2}$ state. 
The corresponding term values are listed in Table~\ref{tab:data-d-state}. 
The ${n=\text{\numlist{50;60}}}$ states were excited via the intermediate $\nSLJF{5s5p}{3}{P}{1}{9/2}$ state, allowing the creation of states with ${F=\text{\numlist{7/2;9/2;11/2}}}$. 
The ${n=\text{\numlist{97;98}}}$ states were excited via the intermediate $\nSLJF{5s5p}{3}{P}{1}{11/2}$ state, allowing the creation of ${F=\text{\numlist{9/2;11/2;13/2}}}$ states. 
Figure~\ref{fig:spectra_D} also includes the best theoretical fit that could be obtained to the data. 
This was realized by first determining the values of the quantum defects $\mu^{(0)}_{n,S,L,J}$ that best reproduce the measured energy levels and then using these to update the Rydberg-Ritz expression [Eq.~(\ref{eq:Rydberg-Ritz})] for the $n$ dependence of the quantum defect at {high-$n$} (see Table~\ref{tab:delta}).
The predicted levels shown in Fig.~\ref{fig:spectra_D} are derived using the updated Rydberg-Ritz formulae. 
However, since the measured quantum defects of \Sr{88} $\SLJ{1}{D}{2}$ states (with ${I=0}$) are available up to ${n=70}$, the Rydberg-Ritz expression from \cite{vail12} is used for these states. 
The measured quantum defects for the $\SLJ{3}{D}{}$ states are shown in Fig.~\ref{fig:Rydberg-Ritz-D} together with the values given by both the present and the earlier Rydberg-Ritz expressions. 
The differences between the predicted quantum defects [Eq.~(\ref{eq:Rydberg-Ritz})] based on the present data for \Sr{87} and previous data for \Sr{88} \cite{vail12} appear to be small \num{\sim 0.02}. 
However, when converted to energy, this small difference translates into discrepancies of \SI{130}{\MHz} for ${n=100}$ and \SI{1}{\GHz} for ${n=50}$ well outside the uncertainty of the current experiments. 

The present Rydberg-Ritz formulae can also be tested against earlier measured quantum defects for $\SLJ{}{D}{}$ states in \Sr{87} (${n > 100}$) \cite{sun89}. 
The data are reproduced to within an average difference of \SI{\sim 60}{\MHz}. 
When the modified ionization limit discussed above is used to evaluate the quantum defect, the average difference is reduced to \SI{\sim 25}{\MHz}. 
These residual differences could be caused by stray fields present in the heat pipe used for the earlier work. 
Additionally, the current theoretical model can predict the hyperfine structure of $\SLJ{}{D}{}$-states around ${n \simeq 280}$ which can again be compared with the earlier measurements \cite{ye13}. 
Due to the uncertainty in the ionization threshold, the exact energies cannot be evaluated but the size of the hyperfine splittings is well reproduced within an error of \SI{10}{\MHz}.

Finally, the improved Rydberg-Ritz formulae for the $\SLJ{3}{D}{}$ states determined from the present data for \Sr{87} can be used to determine spectroscopic information for \Sr{88}. 
When we compare energies for the $5s50d$ and $5s80d$ $\SLJ{3}{D}{1,2}$ states derived using the present updated Rydberg-Ritz formulae with earlier measurements \cite{mill11, brid16} the agreement is significantly improved over that obtained using the earlier Rydberg-Ritz parameterization, the differences between theory and experiment being reduced by several hundred \si{\MHz}. 

As a further test of the present theoretical approach, Table~\ref{tab:data-d-state} includes the frequency separations between selected pairs of levels that could be measured during a single FSR scan of the \SI{319}{\nm} laser and that are known to high precision. 
Table~\ref{tab:data-d-state} also includes the corresponding theoretical predictions. 
In all but one case the measured and theoretical separations agree to better than \SI{\pm 10}{\MHz}. 

\begin{table}[!htbp]
\caption{\label{tab:delta}Values of the parameters $\mu_0$, $\alpha$, and $\beta$ for the Rydberg-Ritz formula obtained in this and earlier work.}
\begin{ruledtabular}
\begin{tabular}{cccccc}
Series				& Term				& $\mu_0$			& $\alpha$			& $\beta$			& Reference		\\ \hline
$\nSLJ{5sns}{}{}{}$	& $\SLJ{1}{S}{0}$	& \num{3.26896(2)}	& \num{-0.138(7)}	& \num{0.9(6)}		& \cite{vail12}	\\
$\nSLJ{5sns}{}{}{}$	& $\SLJ{3}{S}{1}$	& \num{3.37065}		& \num{0.443}		& \num{-0.553}		& This work		\\
					&					& \num{3.371(2)}	& \num{0.5(2)}		& \num{-1(2)e1}		& \cite{vail12}	\\ \hline
$\nSLJ{5snd}{}{}{}$	& $\SLJ{1}{D}{2}$	& \num{2.3807(2)}	& \num{-39.41(6)}	& \num{-109(2)e1}	& \cite{vail12}	\\
$\nSLJ{5snd}{}{}{}$	& $\SLJ{3}{D}{1}$	& \num{2.673}		& \num{-5.4}		& \num{-8166}		& This work		\\
					&					& \num{2.658(6)}	& \num{3(2)}		& \num{-8.8(7)e3}	& \cite{vail12}	\\
$\nSLJ{5snd}{}{}{}$	& $\SLJ{3}{D}{2}$	& \num{2.662}		& \num{-15.4}		& \num{-9804}		& This work		\\
					&					& \num{2.636(5)}	& \num{-1(2)}		& \num{-9.8(9)e3}	& \cite{vail12}	\\
$\nSLJ{5snd}{}{}{}$	& $\SLJ{3}{D}{3}$	& \num{2.612}		& \num{-41.4}		& \num{-15363}		& This work		\\
					&					& \num{2.63(1)}		& \num{-42.3(3)}	& \num{-18(1)e3}	& \cite{vail12}
\end{tabular}
\end{ruledtabular}
\end{table}

\section{Summary}

The present work demonstrates that the energies of {high-$n$} \Sr{87} Rydberg states can be accurately determined by diagonalizing an isotope-rescaled Hamiltonian. 
This Hamiltonian is constructed using spectral information for the bosonic isotope (\Sr{88}) which has vanishing nuclear spin combined with the hyperfine interaction present in \Sr{87}. 
The present approach can be implemented for fermionic atoms whenever the energy levels for an isotope with vanishing nuclear spin are available. 
The method can also be applied in reverse allowing determination of spectroscopic information, in particular quantum defects, for bosonic isotopes from the hyperfine-resolved spectrum of the fermionic isotope. 
The major limitation on the accuracy of the present analysis is the uncertainty in the hyperfine-resolved ionization threshold. 
This uncertainty can be removed by focusing on energy differences to a reference level whereupon accuracies of the order of a few \si{\MHz} can be achieved. 

\begin{acknowledgments}
Research supported by the AFOSR (FA9550-17-1-0366), the NSF (1600059), the Robert A. Welch Foundation (C-0734 and C-1844), and the FWF (Austria) (FWF-SFB041 ViCoM, and FWF-SFB049 NextLite). The Vienna scientific cluster was used for the calculations. We thank Ya-Ting Chang, Danyel Cavazos, and Randall G. Hulet for use of their equipment in calibrating our wavemeter.
\end{acknowledgments}

\appendix
\section{Matrix elements of the hyperfine operator $V_{\rm{HF}}$}
\label{app:d-state-eqs}

The matrix elements of the hyperfine operator $V_{\rm{HF}}$ can be evaluated analytically \cite{luri62} and they are listed in the following. 
For the diagonal elements of ${J=2}$ states we find
\begin{eqnarray}
&&\bra{((5sn'd) \, \SLJ{1}{D}{2}, I) F} V_{\rm HF} \ket{((5snd) \, \SLJ{1}{D}{2}, I) F} \nonumber \\
&& \qquad\qquad = - a_{\rm 5s} \lambda K \cos(\theta+\xi) \sin\theta  \delta_{n,n'} \nonumber \\
&& \bra{((5sn'd) \, \SLJ{3}{D}{2}, I) F} V_{\rm HF} \ket{((5snd) \, \SLJ{3}{D}{2}, I) F} \nonumber \\
&& \qquad\qquad = a_{\rm 5s} \lambda K \sin(\theta+\xi) \cos\theta \delta_{n,n'} 
\end{eqnarray}
with ${K = F(F+1)-J(J+1)-I(I+1)}$, ${\lambda = (2\ell+1)/(4\ell(\ell+1))}$, ${\xi=\arcsin(1/(2\ell+1))}$ and ${\ell = 2}$. 
The diagonal elements of ${J=1,3}$ states are
\begin{eqnarray}
&& \bra{((5sn'd) \, \SLJ{3}{D}{1}, I) F} V_{\rm HF} \ket{((5snd) \, \SLJ{3}{D}{1}, I) F} \nonumber \\
&& \qquad\qquad = - \frac{1}{4 \ell} a_{\rm 5s} K \delta_{n,n'} \nonumber \\
&& \bra{((5sn'd) \, \SLJ{3}{D}{3}, I) F} V_{\rm HF} \ket{((5snd) \, \SLJ{3}{D}{3}, I) F} \nonumber \\
&& \qquad\qquad = \frac{1}{4 (\ell+1)} a_{\rm 5s} K \delta_{n,n'} \, . \nonumber \\
\end{eqnarray}
The off-diagonal elements between states with the same ${J=2}$ are
\begin{eqnarray}
&&\bra{((5sn'd) \, \SLJ{1}{D}{2}, I) F} V_{\rm HF} \ket{((5snd) \, \SLJ{3}{D}{2}, I) F} \nonumber \\
&& \qquad\qquad = - \frac{\lambda}{2} a_{\rm 5s} K \cos(2\theta+\xi)  O_{n,n'} \, .
\end{eqnarray}
and those with different $J$ are
\begin{eqnarray}
&& \bra{((5sn'd) \, \SLJ{1}{D}{2}, I) F} V_{\rm HF} \ket{((5snd) \, \SLJ{3}{D}{1}, I) F} \nonumber \\
&& \qquad\qquad = - \frac{1}{4 \ell} a_{\rm 5s} K_- \sin(\theta-\eta) O_{n,n'} \nonumber \\
&& \bra{((5sn'd) \, \SLJ{1}{D}{2}, I) F} V_{\rm HF} \ket{((5snd) \, \SLJ{3}{D}{3}, I) F} \nonumber \\
&& \qquad\qquad = \frac{1}{4 (\ell+1)} a_{\rm 5s} K_+ \cos(\theta-\eta) O_{n,n'} \nonumber \\
&& \bra{((5sn'd) \, \SLJ{3}{D}{2}, I) F} V_{\rm HF} \ket{((5snd) \, \SLJ{3}{D}{1}, I) F} \nonumber \\
&& \qquad\qquad = \frac{1}{4 \ell} a_{\rm 5s} K_- \cos(\theta - \eta) O_{n,n'} \nonumber \\
&& \bra{((5sn'd) \, \SLJ{3}{D}{2}, I) F} V_{\rm HF} \ket{((5snd) \, \SLJ{3}{D}{3}, I) F} \nonumber \\
&& \qquad\qquad = \frac{1}{4(\ell+1)} a_{\rm 5s} K_+ \sin(\theta - \eta) O_{n,n'} \nonumber \\
&& \bra{((5sn'd) \, \SLJ{3}{D}{1}, I) F} V_{\rm HF} \ket{((5snd) \, \SLJ{3}{D}{3}, I) F} \nonumber \\
&& \qquad\qquad = 0
\end{eqnarray}
with ${\eta=\arcsin\sqrt{\ell/(2\ell+1)}}$, ${K_- = \sqrt{(\ell^2-(F-I)^2)((F+I+1)^2-\ell^2)}}$, and ${K_+ = \sqrt{((\ell+1)^2-(F-I)^2)((F+I+1)^2-(\ell+1)^2)}}$. 
Similar to the $\SLJ{}{S}{}$ states, the overlap integral $O_{n,n'}$ of the radial wavefunctions can be evaluated semiclassically \cite{bhat81} and depends only on the effective quantum number, ${n - \mu_{n,S,L,J}^{(0)}}$. 

\bibliography{fermion-rydberg-spectroscopy}

\end{document}